\documentclass[aps,prd,amsmath,amssymb,superscriptaddress,preprintnumbers,preprint]{revtex4-1}

\pdfoutput=1

\usepackage{graphicx}
\usepackage{amsmath}
\usepackage{amssymb}
\usepackage{amsfonts}
\usepackage{physics}
\usepackage{multirow}
\usepackage{textcomp}

\usepackage[dvipsnames]{xcolor}
\usepackage[linktoc=none]{hyperref}
\usepackage[utf8]{inputenc}
\usepackage{braket}
\newcommand{\SD}{\mathcal{S}_{q}}
\newcommand{\mSD}{m_{\SD}}
    
\newcommand{\SE}{\mathcal{S}_{\ell}}
\newcommand{\FE}{\mathcal{F}_{E}}
\newcommand{\FN}{\mathcal{F}_{N}}
\newcommand{\Fl}{\mathcal{F}_{L}}

\newcommand{\FLu}{\mathcal{F}_{L}^\uparrow}
\newcommand{\FLd}{\mathcal{F}_{L}^\downarrow}
\newcommand{\kup}{k_\uparrow}
\newcommand{\kdown}{k_\downarrow}
\newcommand{\F}{\mathcal{F}}
\newcommand{\PW}{\mathcal{W}}
\newcommand{\mSE}{m_{\SE}}
\newcommand{\mF}{m_{\F}}
\newcommand{\mFLu}{m_{\FLu}}
\newcommand{\mFLd}{m_{\FLd}}

\newcommand{\Pl}{\it{l}}
\newcommand{\Pqt}{\it{t}}
\newcommand{\Pqc}{\it{c}}
\newcommand{\Pqu}{\it{u}}
\newcommand{\Pqd}{\it{d}}
\newcommand{\Pqs}{\it{s}}
\newcommand{\Pqb}{\it{b}}
\newcommand{\Pq}{\it{q}}
\newcommand{\Paq}{\it{\bar{q}}}
\newcommand{\Paqc}{\it{\bar{c}}}

\newcommand{\Paqu}{\it{\bar{u}}}

\newcommand{\PSqt}{\it{\tilde{t}}}
\newcommand{\PSqq}{\it{\tilde{q}}}
\newcommand{\Pp}{\mathrm{p}}
\newcommand{\RKs}{\ensuremath{\mathrm{R_{K^*}}}}

\newcommand{\RK}{\ensuremath{\mathrm{R_{K}}}}

\newcommand{\RKps}{\ensuremath{\mathrm{R_{K^{(*)}}}}}
\newcommand{\RDps}{\ensuremath{\mathrm{R_{D^{(*)}}}}}

\hypersetup{colorlinks,linkcolor={blue},citecolor={teal},urlcolor={violet}}

\newcommand{\documenttitle}{\textcolor{BrickRed}{The Radiative Flavor Template at the LHC:\\ g-2 and W-mass}}

\newcommand{\INFN}{INFN - Sezione di Napoli, Complesso Univ. Monte S. Angelo, I-80126 Napoli, Italy}
\newcommand{\UNINA}{Dipartimento di Fisica ``Ettore Pancini'', Università degli studi di Napoli ``Federico II'', Complesso Univ. Monte S. Angelo, I-80126 Napoli, Italy}

\begin{document}

\title{\documenttitle}
\author{Giacomo Cacciapaglia}
\email{g.cacciapaglia@ip2i.in2p3.fr}
\affiliation{Institut de Physique des 2 Infinis de Lyon (IP2I), UMR5822, CNRS/IN2P3,  F-69622 Villeurbanne Cedex, France}
\affiliation{University of Lyon, Universit\'e Claude Bernard Lyon 1, F-69001 Lyon, France}
\author{Antimo Cagnotta}
\email{antimo.cagnotta@unina.it}
\affiliation{\UNINA}
\affiliation{\INFN}
\author{Roberta Calabrese}
\email{rcalabrese@na.infn.it}
\affiliation{\UNINA}
\affiliation{\INFN}
\author{Francesco Carnevali}
\email{francesco.carnevali@unina.it}
\affiliation{\UNINA}
\affiliation{\INFN}
\author{Agostino De Iorio}
\email{agostino.deiorio@unina.it}
\affiliation{\UNINA}
\affiliation{\INFN}
\author{Alberto Orso Maria Iorio}
\email{albertoorsomaria.iorio@unina.it}
\affiliation{\UNINA}
\affiliation{\INFN}
\author{Stefano Morisi}
\email{smorisi@na.infn.it}
\affiliation{\UNINA}
\affiliation{\INFN}
\author{Francesco Sannino}
\email{sannino@cp3.sdu.dk}
\affiliation{\UNINA}
\affiliation{\INFN}
\affiliation{Scuola Superiore Meridionale, Largo S. Marcellino, 10, 80138 Napoli NA, Italy}
\affiliation{CP$^3$-Origins and Danish-IAS, Univ. of Southern Denmark, Campusvej 55, 5230 Odense M, Denmark}

\begin{abstract}
The Standard Model of particle physics and its description of nature have been recently challenged  by a series of precision measurements performed via different accelerator machines.
Statistically significant anomalies emerged when measuring the muon magnetic momentum, and very recently when deducing the mass of the $\PW$ boson. Here we consider a radiative extension of the Standard Model devised to be sufficiently versatile to reconcile the various experimental results while further predicting the existence of new bosons and fermions with a mass spectrum in the TeV energy scale. The resulting spectrum is, therefore, within the energy reach of the proton-proton collisions at the LHC experiments at CERN. 

The model investigated here allows us to interpolate between composite and elementary extensions of the Standard Model with an emphasis on a new modified Yukawa sector that is needed to accommodate the anomalies. Focusing on the radiative regime of the model, we introduce interesting search channels of immediate impact for the ATLAS and CMS experimental programs such as the associate production of Standard Model particles with either invisible or long-lived particles.  We further show how to adapt earlier supersymmetry-motivated searchers of new physics to constrain the spectrum and couplings of the new scalars and fermions. Overall, the new physics template simultaneously accounts for the bulk of the observed experimental anomalies while suggesting a wide spectrum of experimental signatures relevant for the current LHC experiments. 
  
\end{abstract}
\maketitle

\section{Introduction}
\label{sec:introduction}
In the first two decades of this millennium, increasing experimental evidence of the existence of new physics (NP) beyond the Standard Model (SM) has been accumulated. Several standard deviations from the SM predictions have been observed when determining the anomalous magnetic moment ($g-2$) of the muon by the E821 experiment \cite{Bennett:2006fi} at Brookhaven National Laboratory, and in the semileptonic decays $\mathrm{B} \to \mathrm{D}^{(*)}l\nu$ \cite{babarlfvd, bellelfvd, lhcblfvd} and $\mathrm{B}\to \mathrm{K}^{(*)}ll$ \cite{babarlfvk} by the BaBar, Belle and LHCb Collaborations. These anomalies have been confirmed by the most recent measurement performed by the Muon $g-2$ Collaboration at Fermilab \cite{Abi:2021gix} and at flavor experiments \cite{bellelfvk, lhcblfvk}. Notably, the anomalies in the semileptonic $B$ meson decays emerge due to the observed differences between the decay rates in different lepton families, encoded in the ratios $\RKps$ and $\RDps$. 
In the third decade of this century the CDF II Collaboration at Fermilab  offered an unexpected sign of new physics by unveiling the  high-precision measurement of the $\PW$ boson mass \cite{doi:10.1126/science.abk1781}  from decade-old Tevatron data. The result is seven standard deviations away from the SM prediction. Even after taking into account higher-order corrections \cite{Isaacson:2022rts} and the proper average with previous measurements \cite{Crivellin:2021sff, DAlise:2022ypp}, which reduce the tension to a little less than 4 standard deviations \cite{DAlise:2022ypp,Cacciapaglia:2022xih}, this result clearly points toward a tension with the SM \cite{Strumia:2022qkt,deBlas:2022hdk,Lu:2022bgw}. The most recent LHCb results~\cite{LHCb:2022zom,LHCb:2022qnv} do, however, show a compatibility of $\RKps$ with the SM within two standard deviations.
All this evidence \cite{DAlise:2022ypp} adds to long-standing indirect hints of NP, most notably the dark matter problem, the origin of neutrino masses that emerge from the observation of neutrino oscillations, and the origin of the baryon asymmetry in the Universe. 

Models with additional fermions and scalars, which couple to a single SM quark or lepton via new Yukawa-like couplings not involving the SM Brout-Englert-Higgs field, are prime candidates to explain the $g-2$ anomalies while taking into account the $\mathrm{B}$ meson data~\cite{Arnan:2019uhr, Arcadi:2021glq, Arcadi:2021cwg, Arnan:2016cpy,Calibbi:2018rzv, Crivellin:2018qmi, Crivellin:2021rbq}. The NP contributions arise at loop level, while additional bounds come, most notably, from $\mathrm{B}_0$--$\overline{\mathrm{B}}_0$ mixing and effects on the $Z$ and Higgs bosons couplings to muons. Such models, depending on the quantum numbers of the new states, can also contain a dark matter candidate~\cite{Calibbi:2018rzv}. A model of this kind has been used in Ref.~\cite{Cacciapaglia:2021gff} as a template interpolating perturbative models and strongly interacting models, like technicolor like ones and models of fundamental partial compositeness~\cite{Sannino:2016sfx,Cacciapaglia:2017cdi,Sannino:2017utc}. In the latter limit, the multiplicity of the new fermions and scalars is due to a new confining gauge interaction. 
The inclusion of the $\RDps$ anomalies in this class of models requires further investigation, as loop effects can hardly compete with the tree-level contribution to the $\mathrm{B} \to \mathrm{D}^{(*)}\ l\ \nu$ decay processes in the SM.
Corrections to the $\PW$ boson mass can also be explained in the context of this model as coming from additional fermion loops, providing additional constraints on the new particles and couplings.

In this paper, we aim to describe the productions and decays of the new families of bosons and fermions in proton-proton collisions, in the conditions akin to the ones provided by the Large Hadron Collider (LHC) and future hadron collider projects. We will, in particular, estimate the limits on the parameter space from current LHC searches, and identify promising new channels that deserve further investigation and dedicated searches.

We focus on the radiative case, where the new fermions and scalars can be directly produced. In the composite case they are confined in new meson and baryon states, that need to be studied at the LHC via their effective interactions.

This article is structured as follows: Sec.~\ref{sec:framework} gives a brief overview of the framework following the notation adopted in Ref.~\cite{Cacciapaglia:2021gff}, establishing the terminology and the assumptions made in the rest of the paper. Section~\ref{sec:lhc_signatures} describes the signatures that could arise in proton-proton collisions, following different hypotheses on the mass hierarchies and on the parameter range allowed by the precision measurement constraints, while Section~\ref{sec:lhc_constraints} focuses on describing in detail the phenomenological implications at the LHC of one specific decay channel via the reinterpretation of one existing search by the Compact Muon Solenoid (CMS) experiment. Finally, Sec.~\ref{sec:results} reviews the constraints on the anomalies in the context of this model and draws conclusions on the potential for future studies at the LHC, before the conclusions in Sec.~\ref{sec:concl}.
\section{Theoretical framework}
\label{sec:framework}

Following Ref.~\cite{Cacciapaglia:2021gff}, the class of models described here naturally interpolates between dynamical models of electroweak symmetry breaking and perturbative models where radiative corrections contribute to flavor observables. In the latter case, the new strong gauge symmetry is demoted to a global one, acting on the new fermions and scalars, hence simply counting their multiplicity. The main contribution to anomalies, therefore, stems from loop diagrams involving the new fermions and scalars. These loops also contribute to the Yukawa couplings of the SM fermions; hence, one can design models where light fermion masses and flavor structures can be radiatively generated \cite{Baker:2021yli}. In this work, we will consider a more general scenario, where direct couplings of the SM fermions to the Higgs boson are also present, and SM fermion masses emerge from a combination of the tree-level couplings and the loop contributions. For concreteness, we will consider a model where the SM is extended by means of a set of Dirac fermions, an $SU(2)_L$ doublet $\Fl$ and two $SU(2)_L$ singlets $\FN$ and $\FE$, and two scalar fields $\SE$ and $\SD$. Their assigned quantum numbers are shown in Table~\ref{tab:1}. We can appreciate that the hypercharge of these new fundamental particles is fixed once the parameter $Y$ is chosen. Integer values of $Y$ for the radiative case are not allowed, because they would lead to fractionally charged elementary states, and $|Y|>1/2$ are not allowed because they would not admit neutral particles.

In this paper we will focus our attention on the  case $Y=1/2$ \cite{Cacciapaglia:2021gff}. 

The NP resides in the Yukawa sector of the theory by means of the most general Yukawa Lagrangian involving the new fermions that is also, by construction, invariant under the global $\mathcal{G}_{\rm TC}$ symmetry. Besides kinetic terms and masses for the new fermions and scalars, the SM Lagrangian is complemented by the following set of Yukawa-like interactions:
\begin{equation}
\begin{split}
    -L_{\text{Yuk},  \text{NP}} =\ & y^{ij}_L\ L^i\mathcal{F}_L\big(\mathcal{S}^j_E\big)^\ast+y^{ij}_E\ \big(E^i\big)^c\mathcal{F}^c_N\mathcal{S}^j_E+\\
    &y^{ij}_Q Q^i\mathcal{F}_L\big(S^j_D\big)^\ast+y^{ij}_D\big(D^i\big)^c\mathcal{F}^c_N \mathcal{S}^j_D+y^{ij}_U\big(U^i\big)^c\mathcal{F}^c_E\mathcal{S}^j_D+\\
    &\sqrt{2}\ \big(\kup\ \mathcal{F}_L\mathcal{F}^c_N+\kdown\ \mathcal{F}_E\mathcal{F}^c_L\big)\Phi_H+\qq{h.c.,} 
\end{split}
\label{eq:lagrangian}
\end{equation}
\noindent
where $\Fl =\left( \FLu ,\FLd \right)^{\text T}$ with respect to $SU(2)_L$ while $Q,\ U^c,\ D^c,\ L,\ E^c$ are the SM fermions expressed in terms of chiral left-handed spinors. Note also that the SM flavor indices $i,j$ are carried by the scalars. The field $\Phi_H$ is the SM Higgs scalar doublet. The complete Lagrangian also contains a generic scalar potential, to complement the usual SM Lagrangian. Also, the fields $R^c$ transform as the representation conjugated to $R$, for example $U^c = (\bar{3},\,1)_{-2/3}$.  
Further terms that violate $\mathcal{G}_{\rm TC}$ can be added to Eq.~\eqref{eq:lagrangian}, and they will be considered in a follow-up work.

For the choice $Y=1/2$ we made in this analysis, the electric charges of the new fermions and scalars are fixed, as reported in  Table~\ref{tab:1}. We see that $\SD$ has the same charge as up-type quarks, while $\FLd$ and $\FE$ are neutral, and $\FLu$ and $\FN$ carry positive charge. This implies, for instance, that $\SD^j$ can decay into an up-type quark plus a neutral heavy fermion or a down-type quark and a heavy charged fermion. This will be further explored in Sec.~\ref{sec:lhc_signatures}, where a detailed description of the experimental signatures will be given. The scalars $\SE^j$ are also neutral: Since, in the absence of $\mathcal{G}_{\rm TC}$-violating operators the new fermions and scalars cannot decay into SM states, we need the lightest state to be neutral and, therefore, act as a potential dark matter candidate.
Having fixed $Y$, the remaining parameters of the model are the Yukawa matrices $y_{Q,U,D}^{ij}$, $y_{L, E}^{ij}$, the couplings $k_{\uparrow, \downarrow}$, and the masses of the new scalars and fermions.
\begin{table}
\begin{center}
\begin{tabular}{c|c|c|c|c|c|c}
  & $\mathcal{G}_{\rm TC}$ & $SU(3)_c$ & $SU(2)_L$ &$U(1)_Y$ & $Q\ (Y=1/2)$ & $Q\ (Y=-1/2)$\\\hline
$\Fl = \left(\begin{array}{c}
     \FLu  \\
      \FLd
\end{array}\right)$ & $\bf F$ & $1$ & $2$ & $Y$&$\left(\begin{array}{c}
     +1 \\
     0
\end{array}\right)$ &$\left(\begin{array}{c}
     0 \\
     -1
\end{array}\right)$ \\ 
$\FN^c$ &  $\bf \bar F$ & $1$ & $1$ & $-Y-1/2$&$-1$& $0$ \\ 
$\FE^c$ &  $\bf \bar F$ & $1$ & $1$ & $-Y+1/2$&0&$1$ \\  \hline 
$\SE^j$ & $\bf F$ & $1$ & $1$ & $Y-1/2$&0 &$-1$\\
$\SD^j$ & $\bf F$ & $3$ & $1$ & $Y+1/6$&$+\frac{2}{3}$ &$-\frac{1}{3}$\\\hline
\end{tabular}
\caption{\em\label{tab:1} Quantum numbers of the additional fermions $\mathcal{F}$ and scalars $\mathcal{S}$ as in Ref.~\cite{Cacciapaglia:2021gff}. Note that $\mathcal{G}_{\rm TC}$ is a global symmetry in our scenario, while $j=1,2,3$ is a family index. The nomenclature follows the association of $\FE$ with the down-component of the doublet, similarly to electrons in the SM (and analogously for $\FN$. Charges similar to the SM ones are obtained for $Y=-1/2$, as shown in the last column. In this paper, however, we will study the case $Y=1/2$, as shown in the second-to-last column.}
\end{center}
\end{table}

\section{Signatures in proton-proton collisions}
\label{sec:lhc_signatures}

\begin{figure}[tbh!]
\centering
\[\vcenter{\hbox{\includegraphics[height=0.35\textwidth]{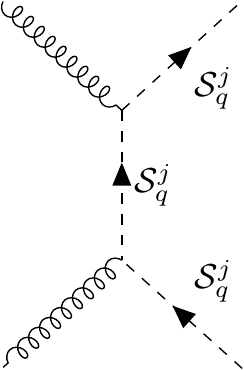}}}\,\,\,\,\,\,\,\,\,
S\vcenter{\hbox{\includegraphics[height =0.25\textwidth]{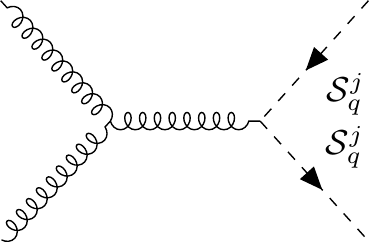}}}
\]
\caption{\em Representative leading-order Feynman diagrams for pair $\SD$ production.}
 \label{fig:production_pair}
\end{figure}

\begin{figure}[tbh!]
\centering
\[\vcenter{\hbox{\includegraphics[height=0.35\textwidth]{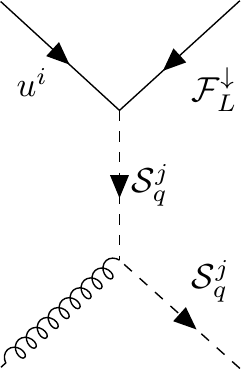}}}\,\,\,\,\,\,\,\,\,
\vcenter{\hbox{\includegraphics[height=0.25\textwidth]{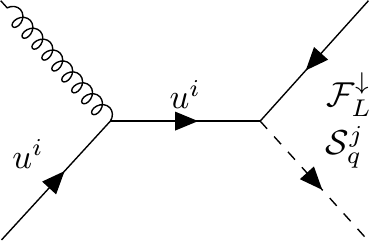}}}
\]
\[\vcenter{\hbox{\includegraphics[height=0.35\textwidth]{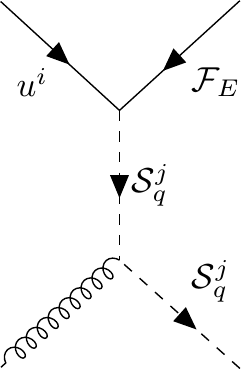}}}\,\,\,\,\,\,\,\,\,
\vcenter{\hbox{\includegraphics[height=0.25\textwidth]{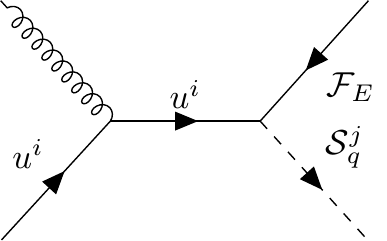}}}
\]
\caption{\em Representative leading-order Feynman diagrams for $\SD$ plus $\F$ production.}
 \label{fig:production_single}
\end{figure}

Constraints on NP scenarios coming from heavy meson physics and electroweak (EW) precision measurement can be complemented by direct searches at the LHC. In our radiative model, the masses of the new fermions and scalars are preferably above the EW scale, hence spanning from a few hundred GeV to a few TeV. As the fermions $\mathcal{F}$ and the scalar $\SE$ are colorless, their production at proton-proton colliders will be very small. On the contrary, the $\SD$ scalars carry color charge; hence, they are produced via gluon and quark fusion, $gg\to\SD\SD^\ast$ and $q\bar{q}\to\SD\SD^\ast$ in proton-proton collisions via QCD interactions. Representative leading-order Feynman diagrams for pair $\SD$ production are shown in Fig.~\ref{fig:production_pair}. 
This kind of production mechanism has a limited dependence on the parameters of the model. Since the color charge and the spin of $\SD$ are set, the pair production cross section only depends on the mass $\mSD$ of the new boson and on the multiplicity: namely $N_{TC}$ and the number of light families.  
Direct production of a single $\SD$ boson, accompanied by one $\Fl$ or $\FE$/$\FN$ fermion, is also possible with representative leading-order Feynman diagrams shown in Fig.~\ref{fig:production_single}. In this case, a NP interaction vertex, e.g., $Q\SD\Fl$, is necessarily involved; hence, the parton-level cross section depends on the details of the models and on the NP couplings. In particular, left-handed production will depend on the coupling $y^{ij}_Q$, while right-handed production will depend on $y^{ij}_U$, $y^{ij}_D$.

This section will cover possible experimental signatures coming from pair production of $\SD$ bosons, and we will focus on the minimal scenario where only one $\SD$ family is present (and all others are heavier). The experimental signatures of $\SD$ boson pairs produced at the LHC depend on the decay chains allowed within the model. The particular realization of the model described in Sec.~\ref{sec:framework} is not fully determined by the precision measurements, which are detailed in Appendix~\ref{sec:prec_constraints}. In fact, it is the mass of the boson $\mSD$, the masses of both up-type and down-type fermions, $\mFLu$ and $\mFLd$, the other fermions $\FE$ and $\FN$ as well as the coupling strengths, that will determine the possible decays for the $\SD$ bosons and the respective branching fractions. 
We describe these final states in two steps: in the first step we address the direct decays of the $\SD$ boson, which depend on the coupling strengths, i.e. $y^{ij}_Q$, $y^{ij}_U$, and $y^{ij}_D$, and on the difference in mass between $\SD$ and the fermions. In the second step we will describe the possible decay cascades and final states that can arise depending on the mass hierarchy between the various fermions $\mathcal{F}$ and $\SE$.
In Sec.~\ref{sec:lhc_constraints}, we will describe the example of one particular realizations that can be used to derive quantitative predictions at the LHC. 

\subsection{Decays of the $\SD$ bosons and the flavor basis}

To properly discuss the phenomenology stemming from $\SD$ pair production, we first need to bring the interactions in Eq.~\eqref{eq:lagrangian} to the mass eigenstate basis. For the scalars, we can assume without loss of generality that the mass matrices are already diagonal; hence, selecting the lightest state consists on fixing the index $j$ in the couplings. In the heavy fermion sector, instead, a mixing is generated by the couplings to the Higgs $\kup$ and $\kdown$. This mixing is proportional to the EW scale versus the vectorlike mass of the $\F$ fermions; hence, it is typically small. In the following, mainly for simplicity, we will neglect mixing effects.

In the quark sector, instead, the flavor mixing cannot be neglected as it plays a crucial role in determining the flavor structure in the $\SD$ coupling. The masses receive a contribution both from a SM-like Yukawa coupling and from loops of the new fermions and scalars, similar to the muon mass sketched in Appendix~\ref{sec:Hmumu}. Hence, the masses to be diagonalized for up- and down-type quarks read:
\begin{eqnarray} 
    M_{u}^{ik} &=& \frac{(y_u^{\rm SM})^{ik} v}{\sqrt{2}} +  N_{\rm TC} \sum_{j=1}^3 \frac{(y_Q^{ij} y_U^{kj})}{16 \pi^2} \frac{\kup v}{\sqrt{2}} \ F_{\rm loop} (M_{\SD}^j, \text{masses})\,, \\
   M_{d}^{ik} &=& \frac{(y_d^{\rm SM})^{ik} v}{\sqrt{2}} +  N_{\rm TC} \sum_{j=1}^3 \frac{(y_Q^{ij} y_D^{kj})}{16 \pi^2} \frac{\kdown v}{\sqrt{2}} \ F_{ \rm loop} (M_{\SD}^j, \text{masses})\,.
\end{eqnarray}
The diagonalization leading to the mass eigenstates, i.e. to the SM quarks, derives from a nontrivial interplay between the SM-like Yukawas and the loop contributions. In general, the flavor structure in the mass matrices is not aligned to the one of the new couplings $y_{Q,U,D}$. In the mass eigenstate basis, the Yukawa couplings $y_{Q,U,D}$ are rotated by unphysical rotation matrices, which cannot be measured independently from the Yukawa couplings themselves. In the following, we consider the matrices in the mass eigenstate basis, without changing the notation for simplicity.
Like in the SM, the only rotation matrix that is physical is the Cabibbo-Kobayashi-Maskawa (CKM) one, deriving from a different rotation of the up and down components of the left-handed fields.
The couplings of $\SD^j$ in the mass eigenstate basis, therefore, read
\begin{eqnarray}
    q_{uL} \FLd (\SD^j)^\ast & \Rightarrow & y_Q^{q j}\;\; q_u=u,c,t\,,\\
    q_{dL} \FLu (\SD^j)^\ast & \Rightarrow & \sum_a V_{qa} y_Q^{a j}\;\; q_d=d,s,b\,,\\
    q_{uR}^c \FE^c \SD^j & \Rightarrow & y_U^{q j}\;\; q_u=u,c,t\,,\\
    q_{dR}^c \FN^c \SD^j & \Rightarrow & y_D^{q j}\;\; q_d=d,s,b\,,
\end{eqnarray}
where we follow the convention that the CKM rotation is assigned to the couplings of the left-handed down-type quarks.

\begin{figure}[h!]
\centering
\includegraphics[width=0.3\textwidth]{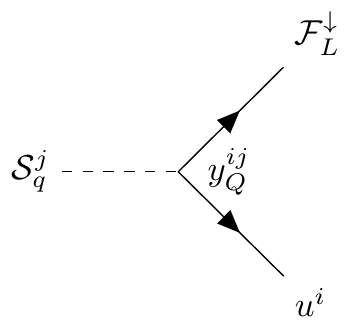}
\includegraphics[width=0.3\textwidth]{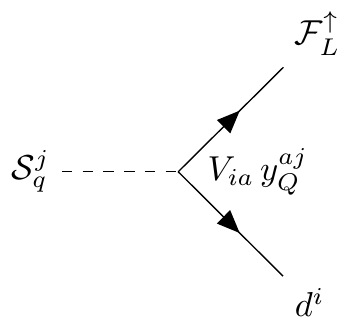}\\
\includegraphics[width=0.3\textwidth]{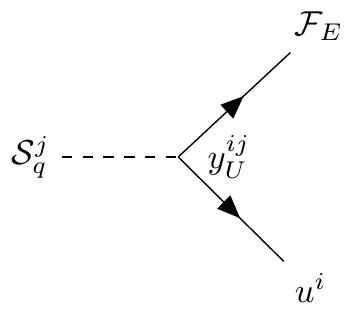}
\includegraphics[width=0.3\textwidth]{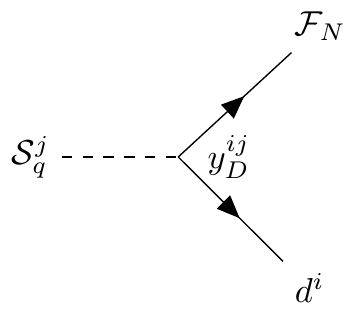}
\caption{\em Decay vertices for $\SD$ bosons to left-handed (top panels) or right-handed (bottom panels) new fermions and an up-type (left panels) or down-type (right panels) quark.}
\label{fig:sd_decays}
\end{figure} 

The first step of the $\SD$ decays is induced by the couplings above, as illustrated in Fig.~\ref{fig:sd_decays}. In the following we will assume that the couplings $y_Q$ are much larger than $y_{U,D}$; hence, we will only consider the decays in the upper diagrams, which involve the doublet fermions $\FLu$ and $\FLd$. The reason for this choice is twofold: On the one side, the products $y_Q y_U$ and $y_Q y_D$ control the loop correction to the quark masses and they need to be small for light quarks (see also discussion for the muon mass); on the other hand, the constraints discussed in Appendix~\ref{sec:prec_constraints} prefer larger contributions in the left-handed sector. Henceforth, the partial decay widths of $\SD$ into up- or down-type quarks are given by
\begin{align}
 \Gamma^j_{q_u}\equiv |y_Q^{qj}|^2  \cdot \frac{\mSD{}_j^2 -(\mFLd-m_{q_u})^2}{32 \pi \cdot \mSD{}_j^3} \cdot (\mSD{}_j^2-\mFLd^2 -m_{q_u}^2)\,, \label{eq:sd_gamma1}\\
  \Gamma^j_{q_d}\equiv \left|\sum_k V_{qk} y_Q^{kj}\right|^2  \cdot \frac{\mSD{}_j^2 -(\mFLu-m_{q_d})^2}{32 \pi \cdot \mSD{}_j^3} \cdot (\mSD{}_j^2-\mFLu^2 -m_{q_d}^2)\,.
\label{eq:sd_gamma2}
\end{align}
Defining the total width of $\SD^j$ as
\begin{equation}
    \Gamma^j = \sum_{q_u = u,c,t} \Gamma^j_{q_u} + \sum_{q_d = d,s,b} \Gamma^j_{q_d}\,,
\end{equation}
the branching ratios into different quark flavors read
\begin{eqnarray}
BR(q_u,\FLd) &=& \frac{\Gamma^j_{q_u}}{\Gamma^j}\qquad q_u = u,c,t\,; \\
BR(q_d,\FLu) &=& \frac{\Gamma^j_{q_d}}{\Gamma^j}\qquad q_d = d,s,b\,.
\end{eqnarray}
More general formulas, including all the mixing patterns, can be obtained in a straightforward way.

\subsection{Chain decays and detector signatures}

The decay cascades of the $\Fl$, $\FE$, and  $\FN$ produced in the decays of the $\SD$ bosons depend on the respective masses and the mass of $\SE$. For simplicity, in the following discussion we will assume that $\FN$ and $\FE$ are heavier than $\Fl$, as they do not appear in the decays following the left-handed coupling dominance. In particular, if $\FLu$ is the lightest particle of the model, it will not further decay and, being charged, it will appear as a long-lived charged particle (LLCP). We are implicitly assuming that it will decay via a suppressed higher-order operator in SM states (leptons) outside the detector, as it cannot be stable on cosmological timescales.
Instead, if $\FLd$ is the lightest particle, being neutral it will result in missing momentum and energy in the detector, that we will henceforth shorten as ``MET''. Note also that the mass hierarchy between $\FLu$ and $\FLd$ plays a crucial role: The heavier one will decay into the lighter via a real or virtual $\PW$ boson. In fact, the two components of the doublet cannot be degenerate. Reproducing the anomaly in the $\PW$ mass measurement requires a mass split of a few tens of GeV. Even if the mass split vanishes at tree level (for $\kup = \kdown$), one is induced by EW loops and it amounts to roughly $166$~MeV in the large mass limit \cite{Cirelli:2005uq}, and it is enough to generate a prompt decay of the heavier into the lightest (plus a charged pion) \cite{Belyaev:2022qnf}.
Finally, the neutral scalars $\SE^j$ couple to $\Fl$ via the coupling $y_L$; hence, if it is lighter than $\FLu$ and $\FLd$, the fermions will decay into it, producing missing energy plus a lepton. We assume that the lightest $\SE^j$ can be stable and produce MET.

\begin{figure}[tbh!]
\centering
\includegraphics[width=0.3\textwidth]{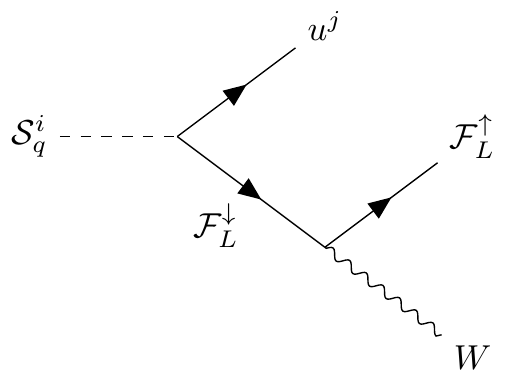}\hspace{0.05\textwidth}
\includegraphics[width=0.3\textwidth]{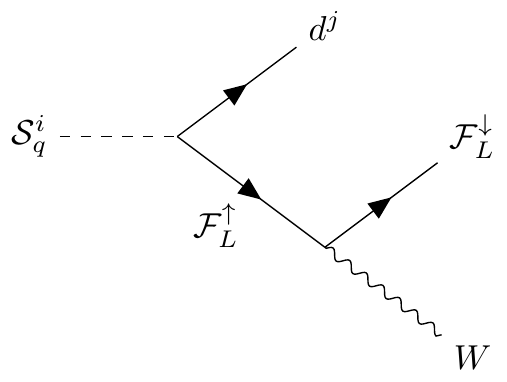}\\
\includegraphics[width=0.3\textwidth]{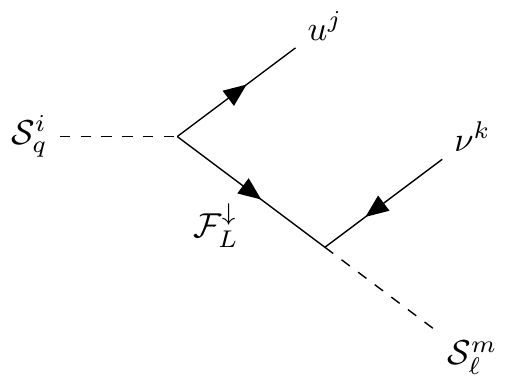}
\includegraphics[width=0.3\textwidth]{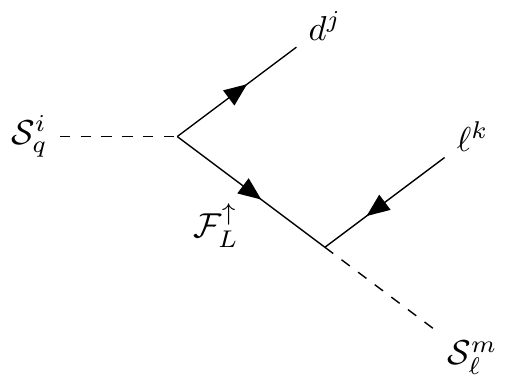}\\
\includegraphics[width=0.3\textwidth]{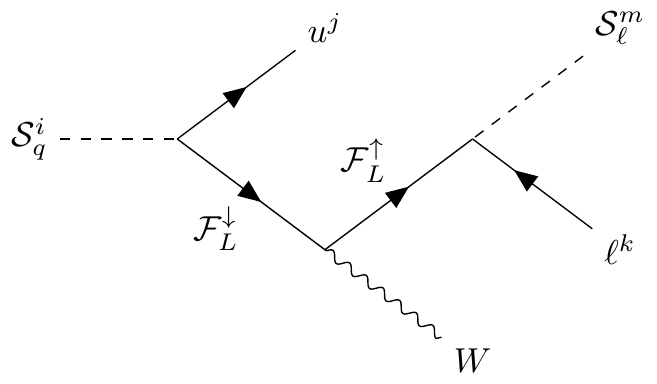} 
\includegraphics[width=0.3\textwidth]{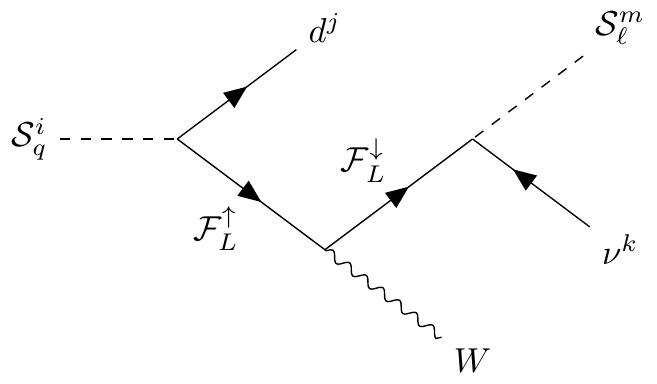}
\caption{\em Decay vertices for $\SD$ bosons to left-handed (top panels) or right-handed (bottom panels) new fermions and an up-type (left panels) or down-type (right panels) quark.}
\label{fig:full_decays}
\end{figure}

The possible decay chains of the fermions $\FLu$ and $\FLd$ are described in the following, depending on the mass ordering, and refer to Feynman diagrams shown in Figs.~\ref{fig:sd_decays} and \ref{fig:full_decays}. We distinguish the following cases:

\begin{enumerate}
    \item $\mSE > \mFLu,\mFLd$: The fermions are the lightest particles, but one fermion will decay into the other plus a $\PW$ boson, either virtual or real. States with either MET or LLCPs are possible, depending on which fermion is the lightest. Representative diagrams for this case are the ones in Figs.~\ref{fig:sd_decays} and~\ref{fig:full_decays}, first row.
    \item $\mSE < \mFLu,\mFLd $: The neutral scalar $\SE$ is the lightest particle, and final states involve missing energy from $\SE$ plus at least one lepton, either charged or neutral, from the NP vertex involving $\SE$ and $\Fl$. Representative diagrams for this case are the ones in Fig.~\ref{fig:full_decays}, second row.
    \item $\mSE < \mFLu,\mFLd $: the scalar $\SE$ is the lightest particle, final states involve missing energy from the $\SE$ plus at least one lepton from from the NP vertex involving the $\SE$ and the $\Fl$ or $\FN$. In this case, one fermion will decay in the other one plus a $W$ boson. Representative diagrams for this case are the ones in Fig.~\ref{fig:full_decays}, third row.
    \item $\mFLd < \mSE < \mFLu $ or $ \mFLu < \mSE < \mFLd $: In this case, one of the fermions can decay into the $\SE$ that further cascades into the second type of fermion plus a lepton, respectively charged or neutral for the two cases.  
\end{enumerate}

It is worthy of note that the difference in mass between the up-type and down-type fermion $|\mFLu-\mFLd|$ needs to be sizeable and in the range of tens of GeV in order to account for the $m_\PW$ measurement, as shown in Appendix~\ref{sec:mw_constraints}. Moreover, $|\mFLu-\mFLd|<m_\PW$ for the largest part of the parameter space, so the $\PW$ boson produced in the decays will be virtual in most of the cases. 
For most parameter choices, $\mSE$ is significantly different from  $\mFLu$ and $\mFLd$; therefore, the scenario 4 from the above list is also unlikely to occur. 

Table~\ref{tab:lhc_decays} summarizes all possible decay chains depending on the mass hierarchy.

\begin{table}[tbh!]
\begin{center}
\resizebox{\textwidth}{!}{
\begin{tabular}{ c |c | c | c |c }
 & $\mSE > \mFLu,\ \mSE > \mFLd$ & $\mSE < \mFLu,\ \mSE < \mFLd$
 & $\mFLd <\mSE < \mFLu$* &  $\mFLu <\mSE < \mFLd$*\\ 
\hline
\multirow{3}{*}{$\mFLu>\mFLd $} & MET+$\Pqu$,& MET+$\Pqu$,& MET+$\Pqu$, &  \\
& MET+$\PW$+$\Pqd$ & MET+$\PW$+$\Pqd$, & MET+$\Pqd$, & NA  \\ 
& & MET+$\Pqd$+$\Pl$ & MET+$\Pqd$+$\Pl$&  \\ 
\hline
\multirow{2}{*}{$\mFLd>\mFLu$} & LLCP+$\Pqd$ & MET+$\Pqu$ & \multirow{2}{*}{N.A.} & MET+$\Pqu$+LLCP+\Pl \\ 
& LLCP+$\Pqd$+$\PW$ & MET+$\PW$+\Pl & & MET+$\Pqu$+$\PW$+LLCP\\ 
 \hline
 \multicolumn{5}{l}{$\Pqu$,$\Pqd$ = up- or down-type quark of the i-th family, $\Pl$ = charged lepton, $\PW$ = $\PW$ boson}\\[-20 pt]
 \multicolumn{5}{l}{LLCP = Long-Lived Charged Particle, MET = missing energy in the detector. }\\[-20 pt]
 \multicolumn{5}{l}{(*) Limited phase space: $\mF \gtrsim 1$~TeV.}\\
 \hline
\end{tabular}
}
\end{center}
\caption{\em \label{tab:lhc_decays} Description of the possible decays at the LHC depending on the mass hierarchy of the particles.}
\end{table}

Finally, a note for the  $Y=-1/2$ scenario: While the diagrams are the same as the $Y=1/2$ case, the quantum numbers of the particles are different, as reported in the last column of Table~\ref{tab:1}. This has the consequence that final states still involve either MET or LLCPs, but in this case, an $\FLu$ contributes to the MET, while an $\FLd$ or an $\SE$ manifests as a LLCP.

\section{Constraints from LHC direct searches}
\label{sec:lhc_constraints}
In order to derive quantitative predictions for observable quantities at the LHC, the model needs to be fully specified, including the coupling structure and the mass hierarchy.
In the following we first describe the scenarios with the minimum number of parameters necessary for explaining the anomalies in muon $g-2$ and $m_\PW$, while accounting for the $\mathrm{B}$ meson sector constraints, then we describe the final states and compare them to existing analyses at the LHC. We focus on the scalars $\SD$, as discussed in the previous section, which have the same quantum numbers as the (right-handed) stop in supersymmetry (SUSY); hence, we will focus on SUSY searches that are very close to our expected signal.

\subsection{Minimal scenarios} 

In the models described in Sec.~\ref{sec:framework}, only the left-handed couplings to quarks are explicitly needed. The minimal benchmark scenario therefore requires both right-handed couplings to quarks being zero. Hence, our ``minimal scenario'' consists of nonvanishing $y_Q^{ij}$ and $y_L^{ij}$, while the other Yukawas are negligible.~\footnote{Alternative scenarios foresee a possible mixing of left- and right-handed components. A second benchmark scenario can be considered with equal couplings to left- and right-handed quarks, i.e.  $y_Q^{ij}=y_U^{ij}=y_D^{ij}$.  Scenarios where the $y_U^{ij}\neq y_D^{ij}=0$ will present a different phenomenology, and different branching ratios for the decays of $\SD$ to right-handed up-type and down-type quarks.}
Furthermore, we will consider only one of the $\SD$ to be light, corresponding to preferential couplings to the third-generation of SM fermions.

In this section, we will consider this scenario where there is only one value for $j$=3, to reflect that the lightest $\SD$ couples preferably to the third-generation quarks. The constraint in the quark sector comes from the coupling between mass eigenstates, referred to as $(y_Qy_Q^\dagger)_{bs}$ in Appendix~\ref{sec:prec_constraints}, which would hint at a preferential coupling to second and third quark generation. A first so-called ``natural'' minimal scenario (NMS), therefore, foresees $y_Q^{13}=0$, $y_Q^{23}\neq0$, and $y_Q^{33}\neq0$. By using the Wolfenstein parametrization for the CKM matrix, it is possible to write the contributions coming from different mass eigenstates as in powers of $\lambda$, and by neglecting $\lambda^4$ terms, one finds:
\begin{align}
(y_Qy_Q^\dagger)_{bs} \equiv \sum_{i,k=1}^3 V_{bi}\  (y_Q)_{i3}\ (y_Q^\dagger)_{3k}\  V^\dagger_{ks}=A\lambda^2 (y_Q^{33})^2  + y_Q^{33}y_Q^{23} + A\lambda^2(y_Q^{23})^2 \,.
\label{eq:yqyqbs_nms}
\end{align}
Another possibility we consider is the ``democratic'' minimal scenario (DMS), where all three couplings are different. By neglecting orders  $\lambda^3$ or higher, one finds:
\begin{align}
(y_Qy_Q^\dagger)_{bs}= A\lambda^2 (y_Q^{33})^2  + y_Q^{33}(y_Q^{23}+\lambda y_Q^{13}) + A\lambda^2(y_Q^{23}+\lambda y_Q^{13})^2\,.
\label{eq:yqyqbs_dms}
\end{align}
It is noteworthy that, in the NMS, decays to first-generation quarks will only be present via CKM mixing, hence strongly suppressed. In the DMS, instead, coupling to first-generation quarks will be present and not necessarily small, and they contribute to the final states to consider when obtaining constraints on the model. In the following we will make use of Eqs~\eqref{eq:yqyqbs_nms} and \eqref{eq:yqyqbs_dms} in order to impose the constraint on $(y_Qy_Q^\dagger)_{bs}$ from the $\mathrm{B}_0 - \overline{\mathrm{B}}_0$ mass mixing and derive the constraint on the branching fractions.

Table~\ref{tab:minimal_scenarios} summarizes the parameters in the considered scenario. 

\begin{table}[tbh!]
\begin{center}
\begin{tabular}{ c|c | c }
 & \multicolumn{2}{|c}{Minimal Scenarios}   \\ 
 \hline
 Parameter & Natural (NMS) &  Democratic (DMS) \\ 
\hline
Number of $\SD$ & $1\times N_{\rm TC}$ & $1\times N_{\rm TC}$  \\ Couplings &   $y_Q^{13}=0$; $y_Q^{23}\neq0$; $y_Q^{33}\neq0$ &     $y_Q^{13}\neq y_Q^{23}\neq y_Q^{33}$\\
$(y_Qy_Q^\dagger)_{bs}$ &  $A\lambda^2 (y_Q^{33})^2  + y_Q^{33}y_Q^{23} + A\lambda^2(y_Q^{23})^2$ & $A\lambda^2 (y_Q^{33})^2  + y_Q^{33}(y_Q^{23}+\lambda y_Q^{13}) + A\lambda^2(y_Q^{23}+\lambda y_Q^{13})^2$  \\ 
\hline
\end{tabular}
\end{center}
\caption{\em \label{tab:minimal_scenarios}
 Description of the minimal scenarios considered.}
\end{table}

\subsection{Squarklike final states: Quark pairs + missing energy}

The particular final states that will be detected at a hadron collider, like the LHC, depend on the decays that are possible, and ultimately on the mass hierarchy of the new particles, as described in Sec.~\ref{sec:lhc_signatures}. If we consider as a benchmark a simple scenario where $\mSE>\mFLu$, $\mSE>\mFLd$ and $\mFLd>\mFLu$, then each $\SD$ can decay either to an up-type quark plus $\FLd$, or a down-type quark plus $\FLu$. The latter further decays to $\FLd$ plus a virtual $\PW$ boson. Hence, the detector signatures for a single $\SD$ decay, as from Table~\ref{tab:lhc_decays}, are a heavy invisible particle, manifesting as a MET, plus either an up-type quark, or a down-type quark and a lepton-neutrino pair or quark-quark pair from the virtual $\PW$ boson. While there are not yet dedicated analyses for this specific model at the LHC experiments, the considered production and decay modes share significant similarities with some SUSY scenarios. In particular, production via strong interaction of an $\SD$ pair, and its decay to an invisible fermion plus a quark, shares several similarities with a pair production of squarks, which then further decay to quarks and neutralinos.
We henceforth make use of  results from the LHC that explore such signatures, in particular Refs.~\cite{CMS:2019zmd, CMS:2021eha}.

For both the NMS and DMS, as at least $y_Q^{33}$ and $y_Q^{23}$ have to be different from zero, both top quarks plus MET and charm quarks plus MET final states need to be studied. The experimental signature for charm and up quarks at the LHC is not distinguishable at the moment, so they both manifest as a jet of hadrons in the detector. 
The observable quantities of interest in this case are the excluded cross sections for the top quark pair plus neutralino production, and the light quark pair plus neutralino production.
It is important to recall that, for the chosen value of the hypercharge, the quantum numbers of the $\SD$ boson are the same as the one of a (right-handed) stop quark, while the quantum numbers of $\FLd$ and $\FLu$ match those of Higgsinos, a neutralino and a chargino, respectively.%

In order to constrain the production cross section of these processes, we perform the reinterpretation of two analyses studying in an exclusive way the production of quark pairs plus missing energy in the final state. 
In Refs.~\cite{CMS:2019zmd, CMS:2021eha} limits on the cross section for $\Pp \Pp \to \PSqt \PSqt^*$, with $\PSqt \to \Pqt \chi_0$ or $\Pqb \chi_+ \to \Pqb \PW \chi_0$ are derived, where following the SUSY notation $\PSqt$ is the stop and $\chi_{0,+}$ are the neutralino and chargino, respectively.
In Ref.~\cite{CMS:2019zmd}, exclusion limits are set on $\Pp\Pp\to\tilde{\Pq}\tilde{\Pq}^*\to \Pq\Paq+\chi_0\chi_0$, with $q$ being a $\Pqu$, $\Pqd$, $\Pqs$, or $\Pqc$ quark. 
The former limit can be used to put constraints on the $\Pp\Pp\to\SD\SD^\ast$, $\SD \to \Pqt\FLd$ or $\SD \to \Pqb\FLu$ production, and the latter for $\Pp\Pp\to\SD\SD^\ast\to \Pqc\Paqc+\FLd\FLd$ or $\Pp\Pp\to\SD\SD^\ast\to \Pqu\Paqu+\FLd\FLd$.%

The inclusive cross sections via strong interaction for $\Pp\Pp\to\SD\SD^\ast$ are identical to the one for $\Pp\Pp\to\PSqt\PSqt^\ast$ multiplied by the number of $\SD$ scalars in the model, i.e. $N_{\rm TC}$. 
In particular, the excluded cross section values are taken as a function of the mass of the stop (or light squark) and neutralino, and interpreted as excluded values of the cross section of an $\SD$ and $\FLd$ with the same masses in the $\FLd +\Pqt$($\FLd+\Pqc$,$\FLd+\Pqu$) final state. In order to constrain the model, different values of $N_{\rm TC}$ as well as the coupling strengths and branching ratios are considered. The excluded cross sections, therefore, are evaluated as follows from the SUSY searches:
\begin{eqnarray}
    \sigma(\SD\SD^\ast, \SD \to \Pqt \FLd \,\mathrm{or}\,\SD \to \Pqb \PW \FLu{}_0) =\sigma(\PSqt\PSqt^*, \PSqt\to \Pqt \chi_0 \,\mathrm{or}\,\PSqt \to \Pqb \PW\chi_0) \times N_{\rm TC}\,,\label{eq:sdsdtt}\\
    \sigma(\SD\SD^\ast)\times (BR(\Pqc,\FLd)+BR(\Pqu,\FLd))^2=\sigma(\PSqq\PSqq^*,\PSqq\to \chi_0) \times N_{\rm TC}\,.\label{eq:sdsdcc}
\end{eqnarray}
We recall that $BR(\Pqu,\FLd)$ is negligible in the NMS, while it could be sizeable in the DMS.

\begin{figure}[htbp] 
    \centering
    \includegraphics[width=0.45\textwidth]{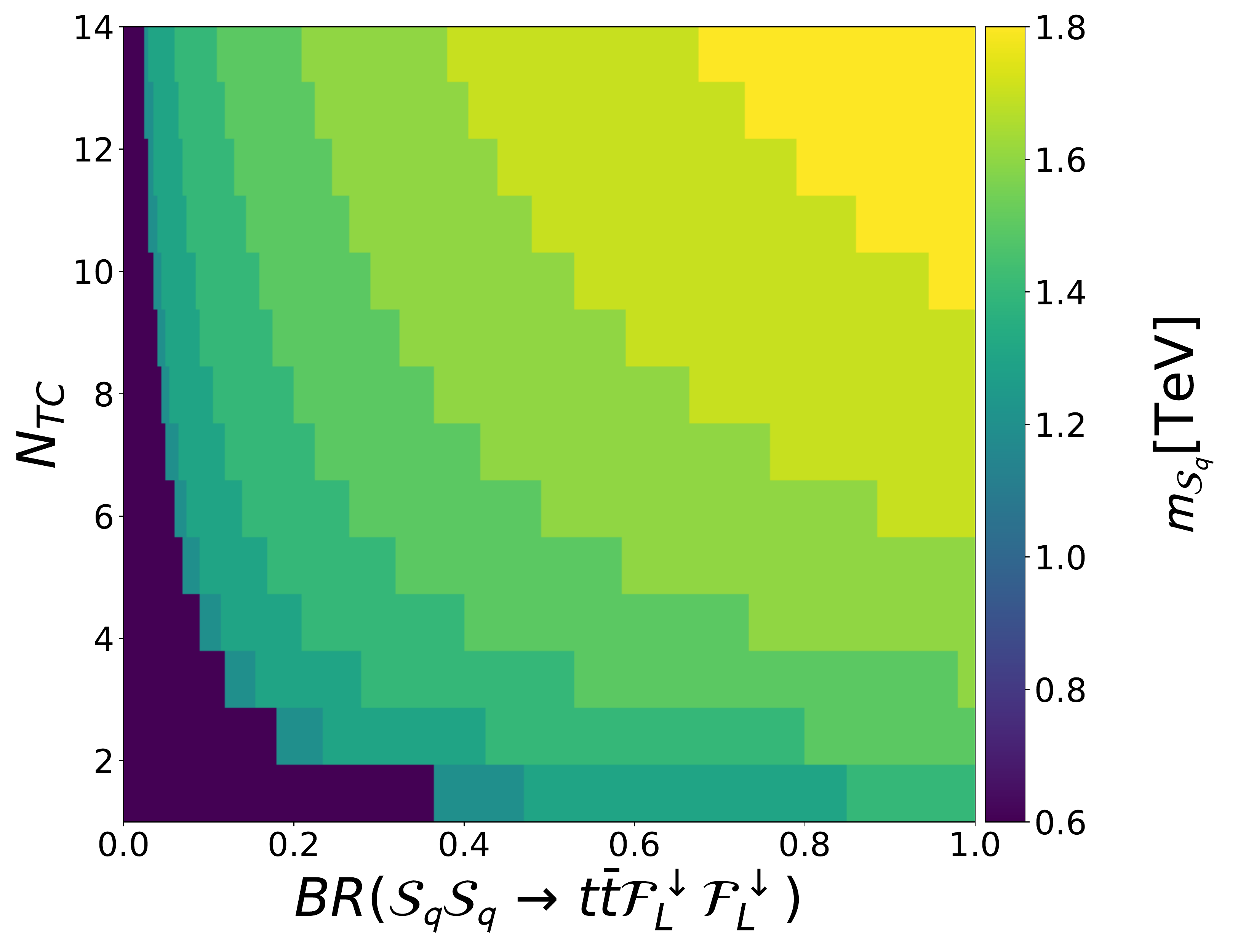}
    \includegraphics[width=0.45\textwidth]{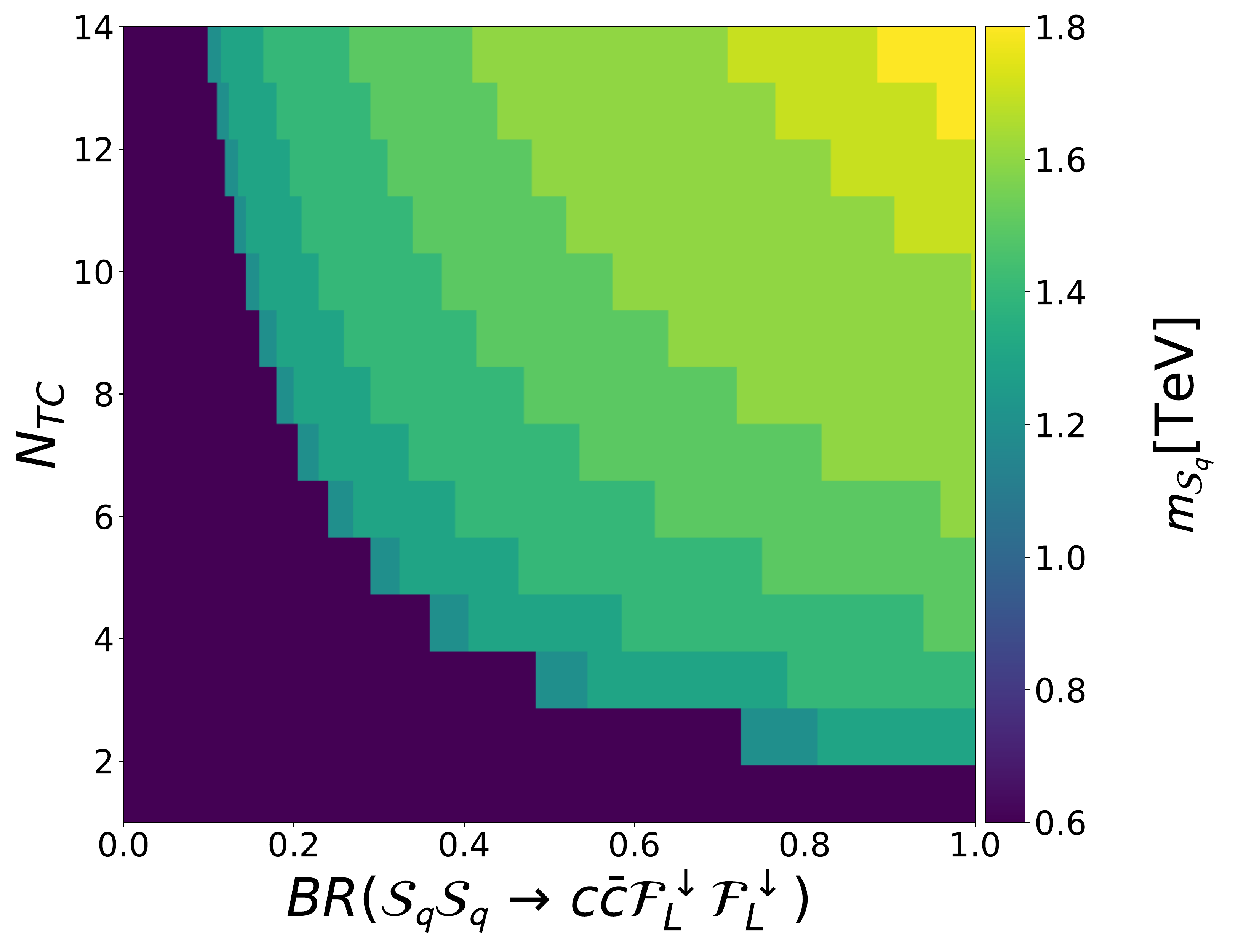}
    \caption{\em Limits on the mass of the new $\SD$ boson vs $N_{\rm TC}$ vs the branching ratio to tt(cc) + $\FLd\FLd$in the tt(cc) + invisible decay channel on the left(right).}
    \label{fig:NTC_BR_MSD_limits}
\end{figure}

\begin{figure}[htbp] 
    \centering
    \includegraphics[width=0.48\textwidth]{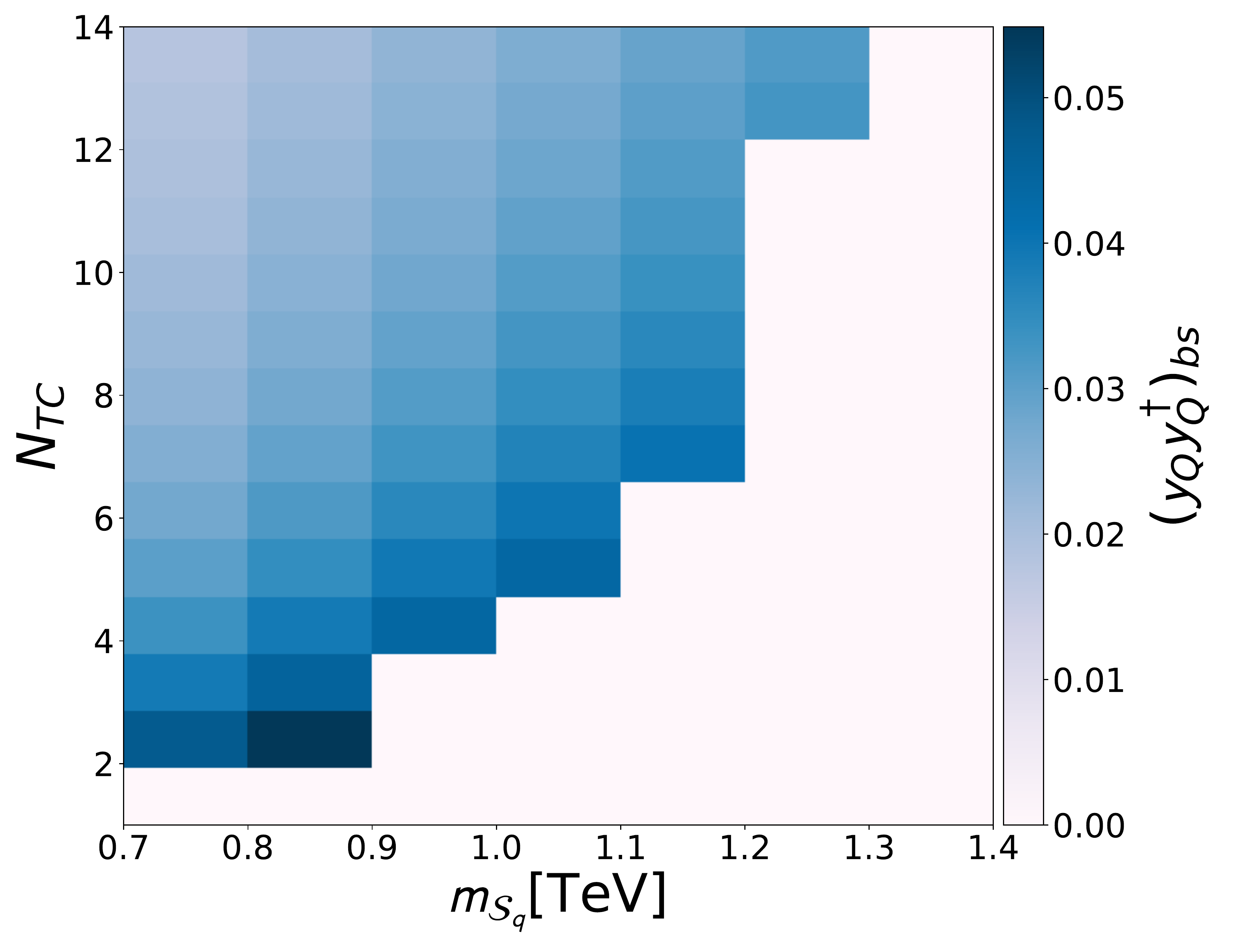}
    \centering
    \includegraphics[width=0.48\textwidth]{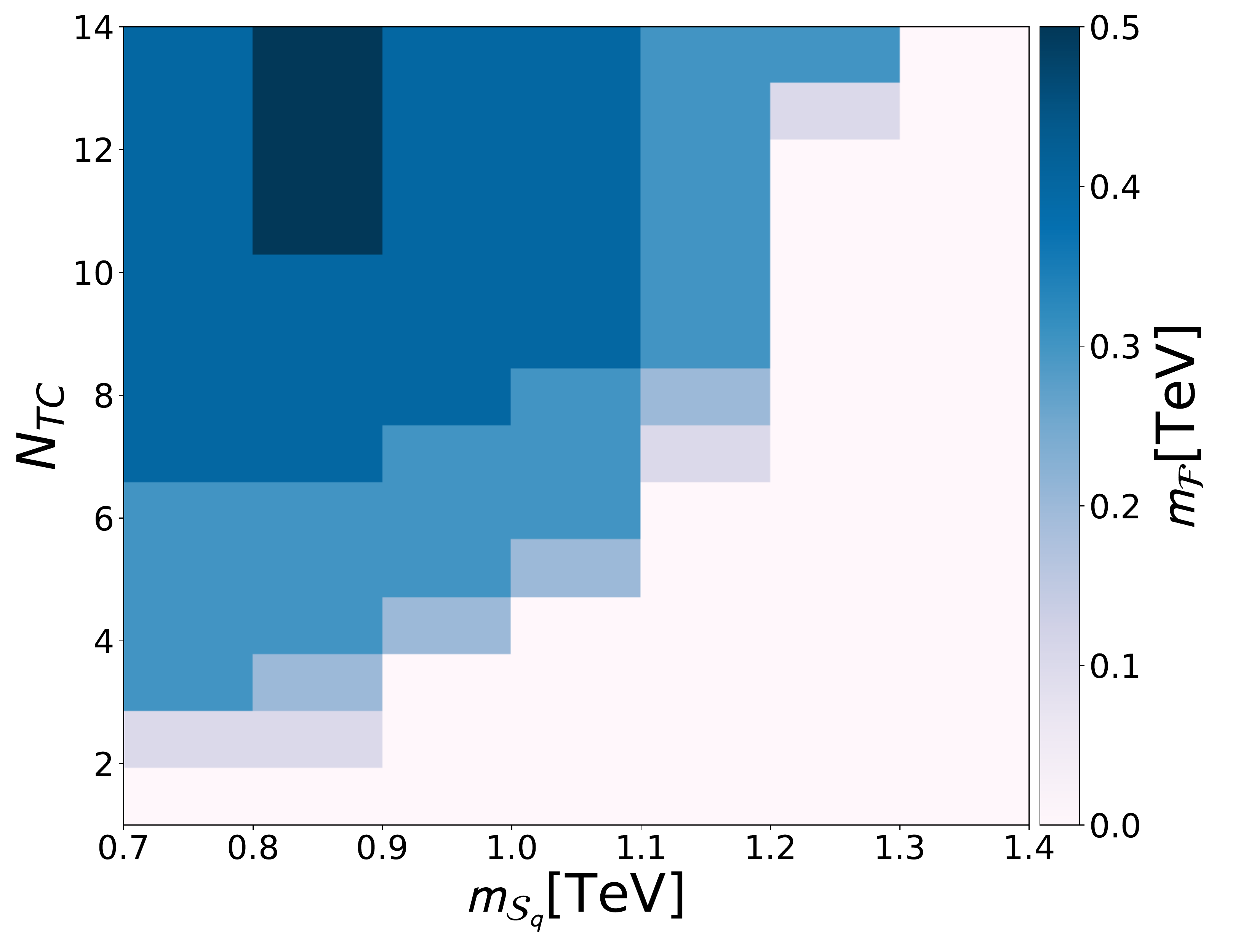}
    \caption{\em  Limits on the mass of the new $\SD$ boson vs $N_{\rm TC}$ and vs the mixed coupling term $(y_Qy_Q^\dagger)_{bs}$ (left, $\mFLd$ = 0.1 TeV), or vs the $\mFLd$ (right).}
    \label{fig:limits_M_NTC}
\end{figure}

The first results that can be extracted are the limits on the $\sigma (\SD\SD) \times BR(\Pqt,\FLd)^2$ and $\sigma (\SD\SD) \times BR(\Pqc,\FLd)^2$, reported in Fig.~\ref{fig:NTC_BR_MSD_limits}.
By use of Eq.~(\ref{eq:sd_gamma1}), it is possible to translate the upper limits on the branching ratio, shown in Fig.~\ref{fig:NTC_BR_MSD_limits}, to a constraint on $y_Q^{33}$, $y_Q^{32}$,  and $y_Q^{31}$. However, in order to extract constraints on the model, the flavor structure needs to be defined. Once $\mSD$ and $\mFLu$ are set, Eqs.~(\ref{eq:Rk}),~(\ref{eq:BB mixing}), and~(\ref{eq:delta a mu}) allow us to identify a unique value of $(y_Qy_Q^\dagger)_{bs}$. By exploiting the relations in Eqs.~(\ref{eq:yqyqbs_nms}) and~(\ref{eq:yqyqbs_dms}), it is possible to identify a range for $y_Q^{33}$, $y_Q^{32}$. If all are excluded, then the model for that combination of $\mSD$, $\mF$, and $N_{\rm TC}$ is excluded. 
Fig.~\ref{fig:limits_M_NTC}, left, shows the limit on $(y_Qy_Q^\dagger)_{bs}$ as a function of $N_{\rm TC}$ and $\mSD$, by keeping $\mFLd$ fixed at 0.1 TeV. Finally, Fig.~\ref{fig:limits_M_NTC}, right, shows the maximum value of $\mFLd$ excluded as a function of $N_{\rm TC}$ and $\mSD$. Those values allow us to strongly constrain the model for values of $\mSD$ in the 0.7 TeV range, but only larger $N_{\rm TC}$ are excluded for values of $\mSD$ above 1 TeV. 

\subsection{Open channels}
This reinterpretation can be extended by performing analyses in different directions: 
\begin{itemize}
    \item The analyses used as a benchmark do not include the potential case where the $\SD$ pair decays to $\Pqt\Pqc+\FLd\FLd$. While this signature might be more challenging, this could help constrain the model further. 
    \item The $\SD$ leg with one LLCP + $\Pqb$, $\Pqs$, or $\Pqd$ quark is not considered. This could lead to challenging but well-identifiable final states with either two LLCP plus two quarks, or one LLCP plus one quark and a top or strange quark and MET. 
    
    \item No single production channel is considered. Further studies are needed to determine whether single production is relevant given the allowed coupling range.
    
    \item Finally, this analysis only refers to the case $\mSE > \FLu$, $\mSE > \FLd$ in Table~\ref{tab:lhc_decays}, further channels are open to study.
\end{itemize}

\section{Summary and results}
\label{sec:results}
\begin{figure}
    \centering
    \includegraphics[width = 0.48 \textwidth]{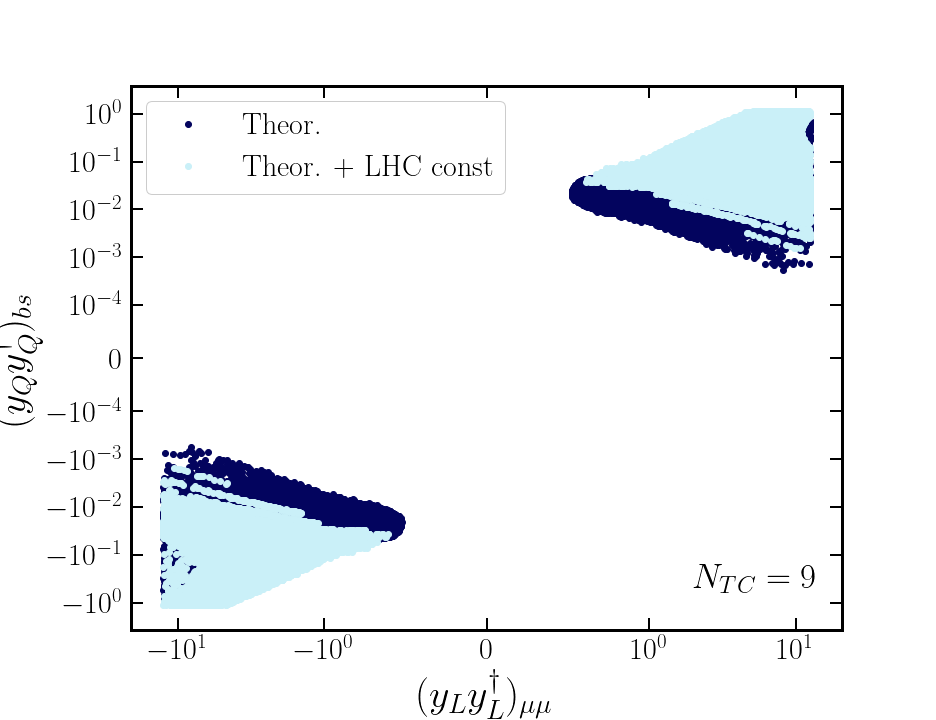}
    \includegraphics[width = 0.48\textwidth]{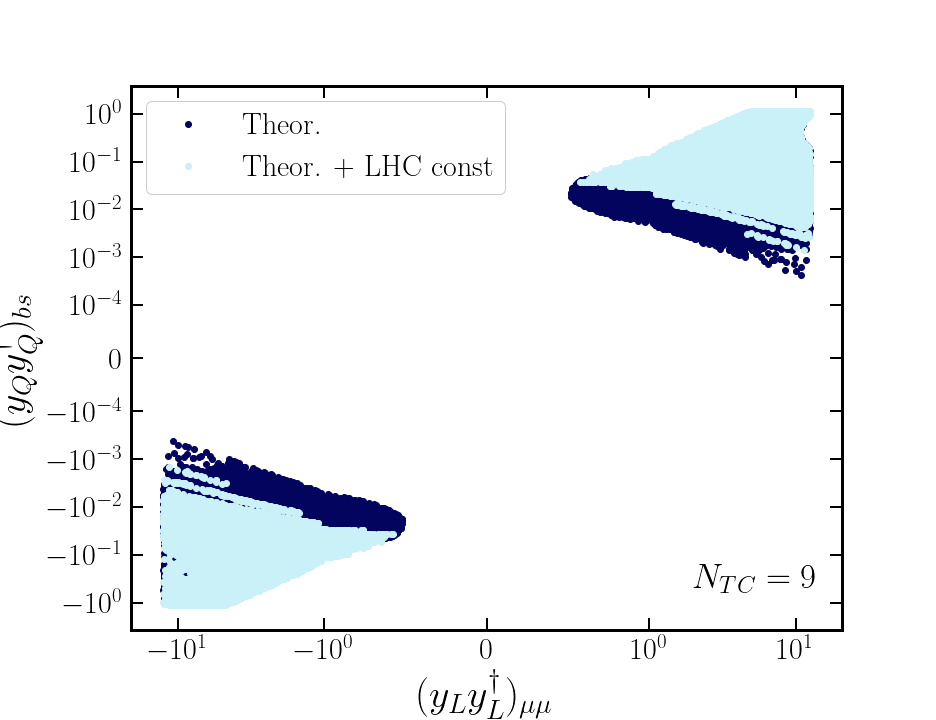}\\
    \caption{\em Allowed values of $(y_Qy_Q^{\dagger})_{bs}$ vs $(y_Ly_L^{\dagger})_{\mu\mu}$ before (dark blue) and after (azure) the application of constraints from the LHC. In the left plot, values of $\Delta a_\mu$ from dispersive measurements~\cite{Abi:2021gix, Aoyama:2012wk,Aoyama:2019ryr,Czarnecki:2002nt,Gnendiger:2013pva,Davier:2017zfy,Keshavarzi:2018mgv,Colangelo:2018mtw,Hoferichter:2019mqg,Davier:2019can,Keshavarzi:2019abf,Kurz:2014wya,Melnikov:2003xd,Masjuan:2017tvw,Colangelo:2017fiz,Hoferichter:2018kwz,Gerardin:2019vio,Bijnens:2019ghy,Colangelo:2019uex,Blum:2019ugy,Colangelo:2014qya, Aoyama:2020ynm} are used, and in the right plot, lattice measurements~\cite{Alexandrou:2022amy} are used.}
    \label{fig:contour_LHC_y}
\end{figure}

\begin{figure}
    \centering
    \includegraphics[width = 0.48 \textwidth]{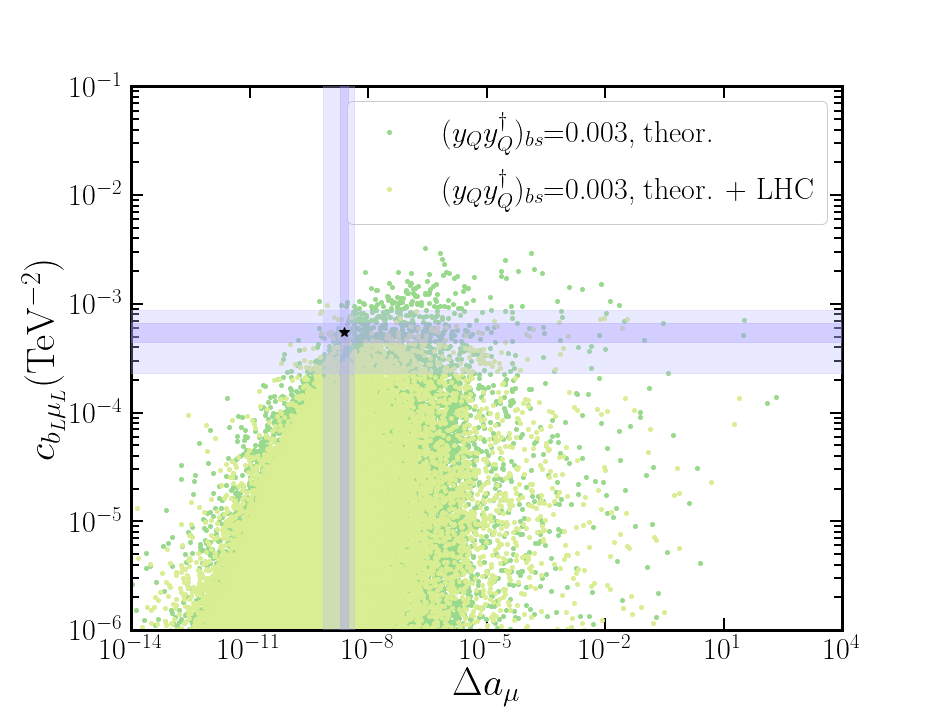}
    \includegraphics[width = 0.48 \textwidth]{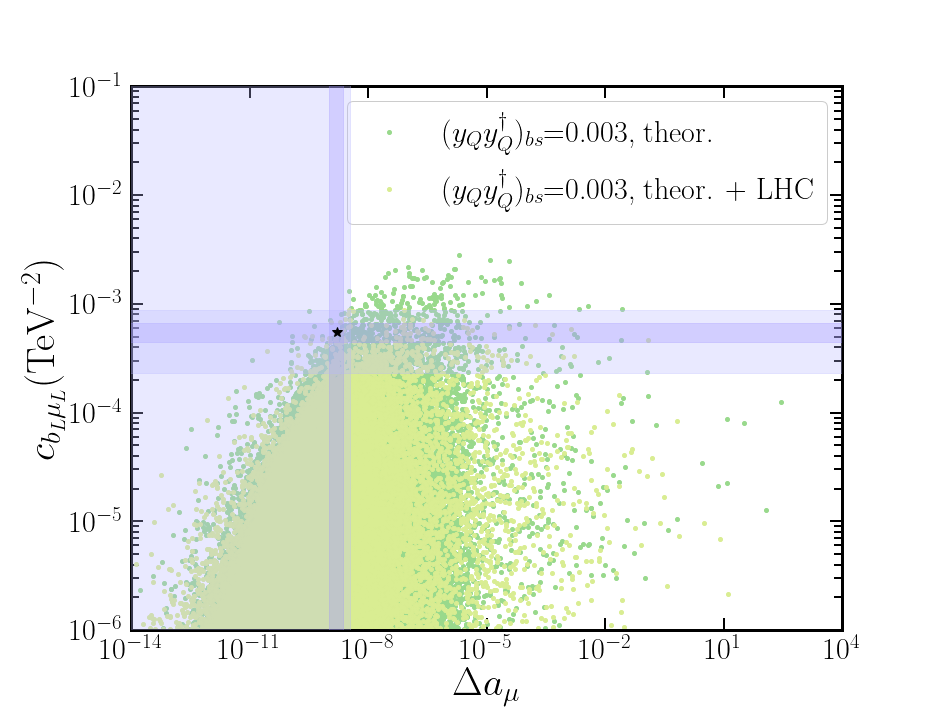}\\
    \caption{\em Allowed values of $c_{b_L \mu_L}$ vs $\Delta a_\mu$ before (dark green) and after (bright green) the application of constraints from the LHC. In the left plot values of $\Delta a_\mu$ from dispersive measurements~\cite{Abi:2021gix, Aoyama:2012wk,Aoyama:2019ryr,Czarnecki:2002nt,Gnendiger:2013pva,Davier:2017zfy,Keshavarzi:2018mgv,Colangelo:2018mtw,Hoferichter:2019mqg,Davier:2019can,Keshavarzi:2019abf,Kurz:2014wya,Melnikov:2003xd,Masjuan:2017tvw,Colangelo:2017fiz,Hoferichter:2018kwz,Gerardin:2019vio,Bijnens:2019ghy,Colangelo:2019uex,Blum:2019ugy,Colangelo:2014qya, Aoyama:2020ynm} are used, in the right plot lattice measurements~\cite{Alexandrou:2022amy} are used. The black stars indicate the best fit for $c_{b_L\mu_L}$ and $\Delta a_\mu$, the dark and light violet lines depict the regions in agreement with the experimental values within 1 and 3 standard deviations, respectively. }
    \label{fig:contour_LHC_c}
\end{figure}

The constraints from the LHC measurements can be used to derive information on the phase space available on $c_{b_L \mu_L}$ and $\Delta a_\mu$ by considering the relations in Eqs.~(\ref{eq:cblmul}) and~(\ref{eq:delta a mu})  when excluding the values of $(y_Qy_Q^{\dagger})_{bs}$ reported in the left panel of  Fig.~\ref{fig:limits_M_NTC} and the values of $\mSD$ and $\mFLd$ reported in the right panel of Fig.~\ref{fig:limits_M_NTC}. 
The limits are evaluated for two different values of $\Delta a_\mu$: First, only dispersive measurements from Refs.~\cite{Abi:2021gix, Aoyama:2012wk,Aoyama:2019ryr,Czarnecki:2002nt,Gnendiger:2013pva,Davier:2017zfy,Keshavarzi:2018mgv,Colangelo:2018mtw,Hoferichter:2019mqg,Davier:2019can,Keshavarzi:2019abf,Kurz:2014wya,Melnikov:2003xd,Masjuan:2017tvw,Colangelo:2017fiz,Hoferichter:2018kwz,Gerardin:2019vio,Bijnens:2019ghy,Colangelo:2019uex,Blum:2019ugy,Colangelo:2014qya, Aoyama:2020ynm} are considered,  then the entire analysis is repeated by taking into account the most recent values from lattice calculations as reported in Ref.~\cite{Alexandrou:2022amy}, and results are reported in the following.
The allowed values of $(y_Qy_Q^{\dagger})_{bs}$ vs $(y_Ly_L^{\dagger})_{\mu\mu}$ before and after the application of LHC constraints for $N_{TC}$ = 9 are reported in Fig.~\ref{fig:contour_LHC_y}. Constraints on $c_{b_L \mu_L}$ and $\Delta a_\mu$, for a representative  value of the coupling of $(y_Qy_Q^{\dagger})_{bs}=0.05$, are reported in Fig.~\ref{fig:contour_LHC_c}. The black star illustrates the best fit for $c_{b_l \mu_L}$ and $\Delta a_\mu$, and the darker and lighter purple bands represent the regions in agreement with the experimental values within 1 and 3 standard deviations, respectively. 

The constraints obtained  hold true for the cases where $|\mFLu-\mFLd|<m_\PW$, and $m_{SE} >\mFLu,\mFLd$. In other scenarios reported in Sec.~\ref{sec:lhc_signatures}, such results only partially constrain the model, as there is freedom for the $\SD$ boson to decay through different decay chains with respect to the ones considered in  Sec.~\ref{sec:lhc_constraints}. Dedicated analyses on proton-proton collision data, searching for LLCPs, or MET plus quarks and leptons, as described in Table~\ref{tab:lhc_decays}, would allow us to detect model's signals for not-yet-excluded scenarios. Even in the absence of a signal, an analysis similar to the one presented in Section~\ref{sec:lhc_signatures} can be applied to further constrain the model. 

\section{Conclusions} \label{sec:concl}

For the past decades several extensions of the SM have been put forward ranging from the time-honored minimal supersymmetric version to composite and radiative realizations including extra dimensions. At the same time, evidence is being accumulated experimentally indicating the existence of new physics beyond the SM. The goal of this work is to establish a phenomenological template  mimicking different models from radiative to composite, depending on how the underlying dynamics is realized, and show that it can also be adopted as an experimental template of new physics by either suggesting new searches or adapting earlier ones to either discover or constrain the parameter space of the theory. Another crucial aspect is that the model is sufficiently flexible to accommodate the various observed SM anomalies. 
  
Specifically, as a first step, we consider a radiative extension of the SM that reconciles the various experimental results while further predicting the existence of new bosons and fermions with a mass spectrum in the TeV energy scale. The resulting spectrum is, therefore, within the reach of the LHC experiments. We suggest new interesting search strategies including the associate production of SM particles with invisible or long-lived charged particles. We further show that it is possible to employ earlier SUSY-inspired searches to constrain the spectrum and couplings of the newly introduced particles. We also check that the template is sufficiently rich to accommodate  the observed experimental anomalies while allowing for new experimental signatures at the LHC.

\subsection*{Acknowledgements}
R.C. and S.M. work was supported by the research grant number 2017W4HA7S ``NAT-NET: Neutrino and Astroparticle Theory Network'' under the program PRIN 2017 funded by the Italian Ministero dell'Universit\`a e della Ricerca (MUR) and by the research project TAsP (Theoretical Astroparticle Physics) funded by the Istituto Nazionale di Fisica Nucleare (INFN).
G.C. is grateful to the LABEX Lyon Institute of Origins (ANR-10-LABX-0066) Lyon for its financial support within the program ``Investissements d'Avenir'' of the French government operated by the National Research Agency (ANR).

\bibliographystyle{utphys}
\bibliography{biblio}

\appendix

\section{Anomalies and high-precision measurement constraints}
\label{sec:prec_constraints}
In this section, we list and review all the precision measurement that significantly constraint the radiative model. Besides the muon anomalies in $\RKps$ and $g-2$, we consider modifications of the Higgs coupling to muons, B$_0-\overline{\mathrm{B}}_0$ mass mixing and the recent discrepancy in the $\PW$ boson mass measurement at CDF II. 
In each subsection we report how the radiative model affects the theoretical estimation of each observable. In the last subsection, we comment how mass splits in the new fermions could explain the $\PW$ boson mass value. The impact of these bounds on collider searches will be discussed in the following section.

\subsection{Radiative muon mass and the Higgs coupling} \label{sec:Hmumu}

The new Yukawa interactions in Eq.\eqref{eq:lagrangian} contribute at one-loop level to all SM fermion masses, in addition to the SM-like Yukawa couplings. In fact, all masses could be generated radiatively, except for the top quark one, whose Yukawa coupling is of order one. In a general setup, cancellations may occur between the tree-level coupling and the loops; hence, one could have sizeable modifications of the Higgs coupling to the fermions.

As we are interested in anomalies in the muon sector, below we illustrate this effect for the muon. 
Schematically, ignoring lepton flavor mixing terms, the muon mass will be given by two competing terms:
\begin{equation} \label{eq:radmuonmass}
    m_\mu = \frac{y_\mu^{\rm SM} v}{\sqrt{2}} + \delta m_\mu^{\rm loop}\,, \qquad  \delta m_\mu^{\rm loop} =  N_{\rm TC} \frac{(y_L y_E)_{\mu\mu}}{16 \pi^2} \frac{\kup v}{\sqrt{2}} \ F_{\rm loop} (\text{masses})\,,
\end{equation}
where $F_{\rm loop}$ is an order one function of the mass parameters \cite{Baker:2021yli} and $m_\mu$ is the measured muon mass. 
The physical coupling of the Higgs boson to muons also receives a correction with respect to the SM value given schematically by
\begin{equation} \label{eq:radmuonhiggs}
    y_\mu^{\rm eff} = y_\mu^{\rm SM}\ \left( 1 + \frac{\delta m_\mu^{\rm loop}}{m_\mu} \ \kup^2\ G_{\rm loop}  (\text{masses})\right)\,, \qquad y_\mu^{\rm SM} =\frac{\sqrt{2} m_\mu}{v}\,, 
\end{equation}
where the function $G_{\rm loop}$ depends on the mass parameters.
Evidence for this coupling has been recently obtained at the LHC Run-2 by the ATLAS and CMS Collaborations \cite{ATLAS:2020fzp,CMS:2020xwi}: Current data are consistent with the SM value at the 43\% level while projections for the full run of the LHC indicate that precision below 5\% is achievable \cite{CMS:2022uuc}.  Hence, the correction encoded in the second term in the parentheses of Eq.\eqref{eq:radmuonhiggs}, proportional to $\delta m_\mu^{\rm loop}$, must be smaller than unity in absolute value.

If we allow for cancellations between the two terms in Eq.\eqref{eq:radmuonmass},  large muon couplings to the new fermions and scalars could generate $\delta m_\mu^{\rm loop} \gg m_\mu$, hence dangerously enhancing the correction to the muon Higgs coupling. There are two ways out: The coupling $\kup$ could be small, however suppressing the radiatively induced SM fermion masses to be of the order of the muon mass for all down-type fermions, or we could ask the product $(y_L y_E)_{\mu\mu}$ to be small for muons. Sizeable effects in muon physics are preserved if $y_L$ is sizeable while $y_E$ is small, as also preferred by $\RK$ (as we will see below). In this work, we will follow the latter possibility. In summary, as long as $(y_L y_E)_{\mu \mu} \sim m_\mu/v$ via a small $y_E$, the correction to the Higgs coupling to muons is under control without affecting the observables in the muon anomalies.

\subsection{\ensuremath{\RK} and \ensuremath{\RKs} anomalies}
\begin{figure}[tbh!]
    \centering
    \includegraphics[width = 0.45 \textwidth]{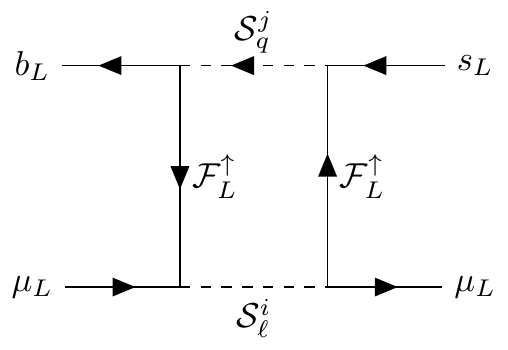}\,\,\,
    \includegraphics[width = 0.45 \textwidth]{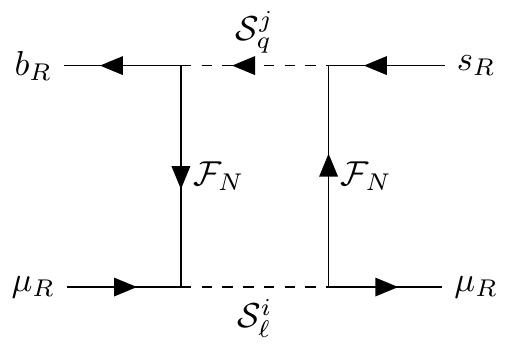}
   \caption{\em \label{fig:Rk} Radiative contribution to the $\mathrm{B} \to \mathrm{K}ll$ transition that affects $\RK$. In the right (left) panel is shown the contribution due to (left) right-handed quarks. The diagram that is used to compute the quantity $c_{b_L\mu_L}$ is the one on the left as $y_E \ll y_L$. }
\end{figure}

The relevant effective Hamiltonian for $\mathrm{B}\to{\rm K}^{(*)}ll$ transitions is given by the following operators \cite{Arnan:2016cpy}:
\begin{equation}
\mathcal{H}^{eff} = - \frac{\alpha}{4\pi} \frac{4G_F}{\sqrt{2}}\  V_{tb}V_{ts}^*\  C_L^{ij}\ \mathcal{O}_L^{ij}\, , 
\end{equation}
where 
\begin{equation}
\mathcal{O}_L^{ij}=\left[\overline{s}\gamma^\mu P_L b\right]\left[\overline{l}^i\gamma_\mu (1-\gamma_5)l^j\right]  \equiv \mathcal{O}_9^{ij} - \mathcal{O}_{10}^{ij}\,.
\end{equation}
We recall that the operator with right-handed quarks is absent in the SM, and it is disfavored as a NP origin of the anomaly by data; hence, we will not consider it here.
The  dimensionful Wilson coefficients are:  
\begin{equation}
    c_{b_L \mu_L} = - \frac{\alpha}{4\pi} \frac{4G_F}{\sqrt{2}} V_{tb}V_{ts}^*\qquad c_L^{\mu \mu} = -\frac{C_L^{\mu\mu}}{(36~{\rm TeV})^2 }
    \label{eq:cblmul}
\end{equation}
that encode NP contributions in the muon final state. More details on the effective Hamiltonian and notation for the coefficients are reviewed in \cite{DAlise:2022ypp}. 
It is possible to rewrite the $\RK$ ratio as follows:
\begin{equation}
\RK=\frac{\left|C_L^{\rm SM}+ \Delta C_L^{\mu \mu}\right|^2}{\left|C_L^{SM}\right|^2}\,,
\end{equation}
where $C_L^{\rm SM} \approx 8.64$ (and we neglect the much smaller contribution to right-handed leptons) \cite{DAmico:2017mtc} is the same for muons and electrons. For $\RKs$ a similar expression holds. 
The one-loop diagrams stemming from the radiative model in Eq.\eqref{eq:lagrangian} are reported in Fig.~\ref{fig:Rk}. As we assume $y_L \gg y_E$ for the muons, the relevant contributions are the one with left-handed quarks. In Ref.~\cite{DAmico:2017mtc} is reported the expression for the Wilson coefficient for the model being examined:
\begin{equation}
    c_{b_L \mu_L} =  N_{\rm TC}  \frac{(y_{L} y_L^\dagger)_{\mu\mu} (y_{Q} y_{Q}^\dagger)_{bs}}{(4\pi)^2 m_{\Fl^\uparrow}^2} \frac{1}{4} F(x,y)
    \label{eq:Rk}
\end{equation}
where $x= m_{\SD}^2/m_{\Fl^\uparrow}^2$, $y= m_{\SE}^2/m_{\Fl^\uparrow}^2$, and the loop functions expression are reported in Appendix\,\ref{App:Loop}. Note that we have neglected the mixing between the fermions $\mathcal{F}$ induced by the $\kup$ coupling in Eq.\eqref{eq:lagrangian}.

To extract bounds on the model parameter space, we impose the latest LHCb result, combining Runs 1 and 2 data, given by \cite{lhcblfvk}
\begin{equation}
\left. \RK\right|_{\rm exp} = 0.846^{+0.042}_{-0.039}(\rm{stat})^{+0.013}_{-0.012}(\rm{syst})\,,
\end{equation}
in the $q^2 = [1.1,6]$~GeV$^{2}$ window. In practice, we require $ 0.807 < \RK < 0.888$ \cite{DAlise:2022ypp}. Note that $\RKs$ leads to similar bounds, and we do not include it for simplicity.

In \cite{DAlise:2022ypp}, the reader can find an in-depth model-independent  study of these anomalies. 

\subsection{\ensuremath{\mathrm{B}_0-\overline{\mathrm{B}}_0}~mass mixing}
\begin{figure}[tbh!]
    \centering
    \includegraphics[width = 0.45 \textwidth]{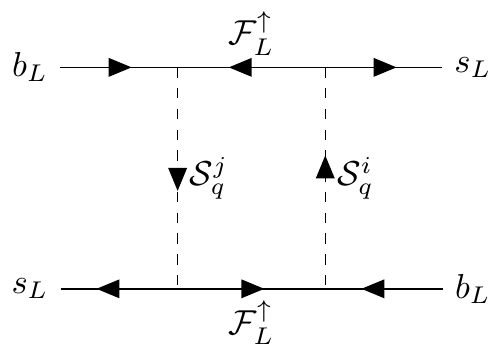}\,\,\,
    \includegraphics[width = 0.45 \textwidth]{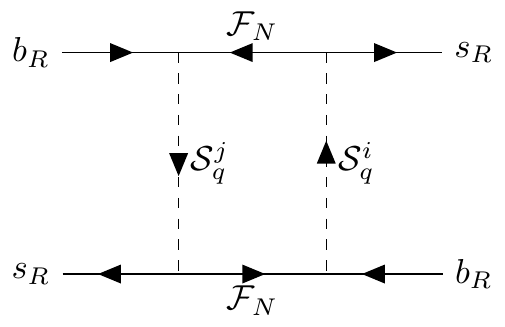}\\
    \vspace{0.5cm}
     \includegraphics[width = 0.45 \textwidth]{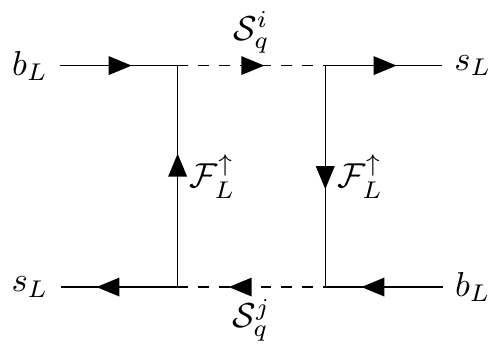}\,\,\,
    \includegraphics[width = 0.45 \textwidth]{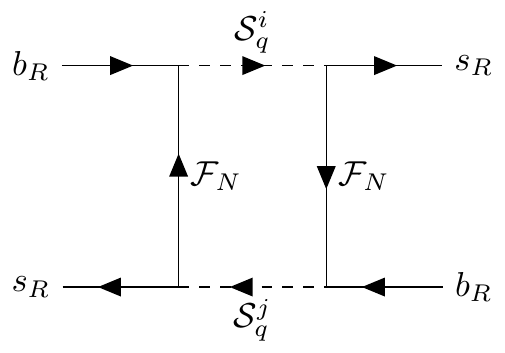}
   \caption{\em \label{fig:BB mixing} Radiative contribution to the $\mathrm{B}-\overline{\mathrm{B}}$ mixing. In the left (right) panel are shown the contributions with left- (right) handed quarks. As for $\RK$, the dominant contribution comes from the operators with left-handed quarks.
   }
\end{figure}

The contributions to the $\mathrm{B}_0-\overline{\mathrm{B}}_0$ mass mixing arise from the diagrams at one-loop level shown in Fig.~\ref{fig:BB mixing}. The dominant contributions are associated with the diagrams with left-handed quarks. These contributions can be written in terms of the following effective Hamiltonian~\cite{Arnan:2016cpy}:
\begin{equation}
\mathcal{H}_{eff}^{{\rm B}\overline{\rm B}}= C_{B\bar B}\left[\overline{s}_\alpha \gamma^\mu P_Lb_\alpha\right]\left[\overline{s}_\beta\gamma_\mu P_Lb_\beta\right].
\end{equation}  
where $\alpha$ and $\beta$ are color indices. The NP contribution from the radiative model under investigation is given by \cite{DAmico:2017mtc}:
\begin{equation}
    C_{B\bar B}^{\rm NP} = N_{\rm TC}  \frac{ (y_{Q} y_{Q}^\dagger)_{bs}^2}{(4\pi)^2 m_{\Fl^\uparrow}^2} \frac{1}{8} F(x,x),
    \label{eq:BB mixing}
\end{equation}
where the loop function is defined in Appendix\,\ref{App:Loop} and $x= m_{\SD}^2/m_{\Fl^\uparrow}^2$.
It is possible to obtain constraints on the Wilson coefficient $ C_{B\bar B}$ in terms of the ratio \cite{Arnan:2016cpy}:
\begin{equation}
R_{\Delta B_s} = \frac{\Delta M_{B_s}^{exp}}{\Delta M_{B_s}^{\rm SM}}-1 = \frac{ C_{B\bar B}^{\rm NP}}{ C_{B\bar B}^{\rm SM}},
\end{equation}
where the SM prediction is $ C_{B\bar B}^{\rm SM}(2\,M_\PW)\simeq 8.2 \times 10^{-5}\,{\rm TeV}^{-2}$ and the current bound on the NP contribution reads $C_{B\bar B}^{\rm NP} \in [-2.8,\ 1.3]\times 10^{-5}\, {\rm TeV}^{-2}$  \cite{Arnan:2016cpy} at 3 standard deviations. 

\subsection{Muon $g-2$ anomaly}
\begin{figure}[tbh!]
    \centering
    \includegraphics[width = 0.4 \textwidth]{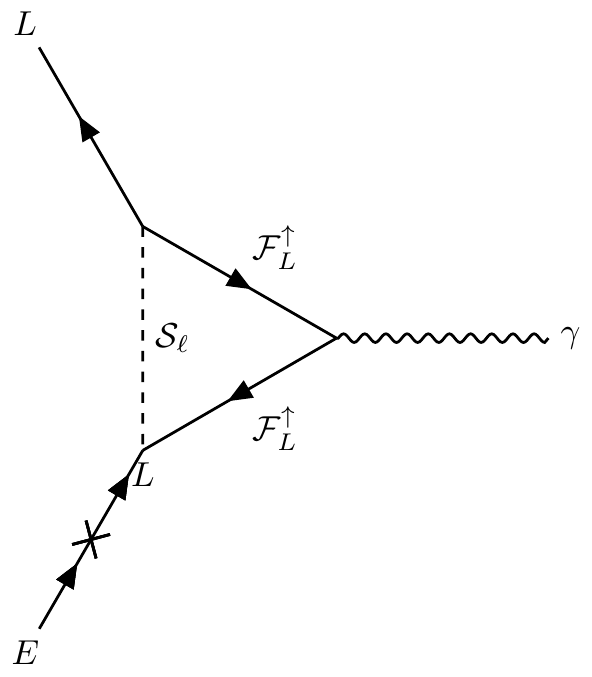}\,\,\,
    \includegraphics[width = 0.4 \textwidth]{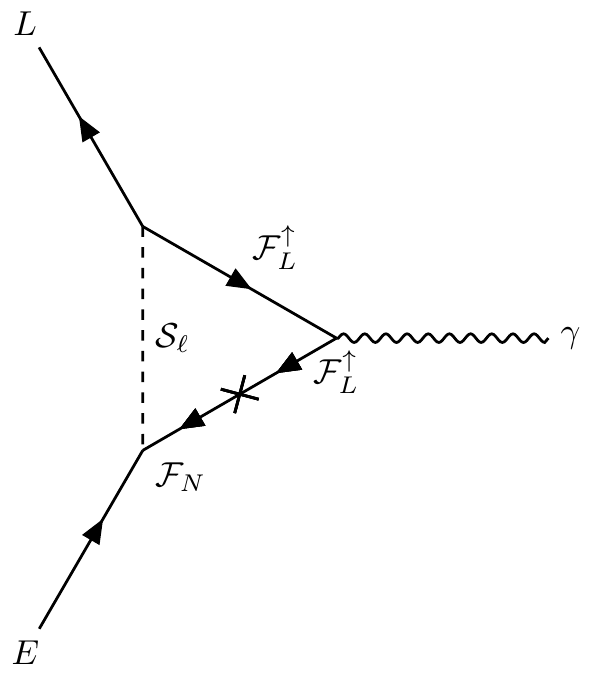}
   \caption{\em \label{fig:g-2} One-loop contributions to  $(g-2)_\mu$ from the radiative model under investigation.}
\end{figure}
The new fermions and scalars  contribute to the anomalous magnetic moment of the muon via their left-handed couplings $y_L$, as shown by the left graph in Fig.~\ref{fig:g-2}. Additionally, the model contains a contribution proportional to the Yukawa coupling $\kup$, as shown by the representative diagram in the right panel of Fig.~\ref{fig:g-2}. This term will be proportional to the product of Yukawas $(y_L y_E)_{\mu \mu} \kup$, which also contributes to the muon mass, as shown in Eq.\eqref{eq:radmuonmass}.
The NP contribution can be computed in terms of the following operator\,\cite{Arnan:2016cpy}:
\begin{equation}
\mathcal{H}_{eff}^{a_\mu} = -a_\mu\frac{e}{4m_\mu}\left[\overline{\mu}\sigma^{\mu\nu}\mu\right]F_{\mu\nu},
\end{equation}
where $m_\mu$ is the muon mass and $e$ is the electron charge. The NP contribution to $a_\mu$ is given by\,\cite{DAmico:2017mtc}:
\begin{multline}
    \Delta a_\mu = \frac{N_{\rm TC}\ m_\mu^2}{(4\pi)^2 m_{\FLu}^2}\left(\frac{(y_{L} y_E)_{\mu\mu}\ \kup v}{m_\mu}\left[2 q_{{\cal S}_E} F_{LR} (y) + 2 q_{{\cal F}}  G_{LR}(y)\right] + \right.\\
    \left.  (y_{L} y_L^\dagger)_{\mu\mu}\left[2 q_{{\cal S}_E}F_7(y) + 2 q_{\cal F}\tilde F_7(y)\right]\right),
    \label{eq:delta a mu}
\end{multline}
where $x= m_{\SD}^2/m_{\FLu}^2$ and $y= m_{\SE}^2/m_{\FLu}^2$. Here, $q_{{\cal F}}$ and $q_{{\cal S}_E}$ are the charges of $\FLu$ ($q_{\mathcal{F}} = -1$) and $\SE$ ($q_{\mathcal{S}_E} = 0$), respectively. In our case, we have only the contribution from $\FLu$ since $\SE$ is neutral. The loop functions are reported in Appendix\,\ref{App:Loop}. 

The main uncertainty on the SM prediction for $a_\mu$ comes from the hadronic contribution to the photon vacuum polarization. Recent lattice results \cite{Borsanyi:2020mff,FermilabLattice:2022smb,Alexandrou:2022amy, Ce:2022kxy} are in tension with data-driven approaches based on dispersion relations \cite{Davier:2017zfy,Keshavarzi:2018mgv,Colangelo:2018mtw,Hoferichter:2019mqg,Davier:2019can,Keshavarzi:2019abf, Colangelo:2022vok}. In the numerical analysis, we consider NP contributions in agreement with the lattice results \cite{Alexandrou:2022amy}, $\Delta a_\mu  = (165\pm 60)\times 10^{-11}$, or dispersion estimates, $\Delta a_\mu  = (251\pm 59)\times 10^{-11}$ \cite{Abi:2021gix}.

\subsection{\ensuremath{M_{\PW}}~anomaly}
\label{sec:mw_constraints}
Recently, the CDF II Collaboration at Fermilab Tevatron Collider reported a new estimation of the $\PW$ boson mass using a sample of approximately 4 $10^6$ $\PW$ bosons. In Ref.~\cite{CDF:2022hxs}, it is reported that
\begin{equation}
    M_{\PW}^{\rm CDF} = 80433.5\pm 6.4_{stat}\pm 6.9_{syst} = 80433.5\pm 9.4~{\rm MeV}.
    \label{eq:CDF_W}
\end{equation}
The theoretical prediction for the same quantity within the SM, as reported in Ref.~\cite{ParticleDataGroup:2020ssz}, is:
\begin{equation}
    M_{\PW}^{\rm th} = 80357\pm 4_{inputs}\pm 4_{theory}~{\rm MeV}.
\end{equation}
The theoretical estimation is obtained with a combination of perturbative expansions and a set of high-precision measurements of observables in the EW sector. The uncertainty on such a quantity can be divided into two components. One is related to the uncertainties of the observables used, and the latter is related to higher-order terms in the perturbative SM calculation. It can be easily shown that
\begin{equation}
    \left.\Delta M_{\PW}\right|_{\rm CDF}= M_{\PW}^{\rm CDF}-M_{\PW}^{\rm th} = 76\pm 11~{\rm MeV}\,.
\end{equation}
This means that there is a 7 standard deviation discrepancy between the CDF experimental measurement and the theoretical estimation. Previous measurements, summarized in Table~\ref{tab:MW}, have a smaller accuracy and central values closer to the SM prediction. A weighted average should be considered, however with error increased by the $\sqrt{\chi^2/\text{ndf}}$ factor \cite{ParticleDataGroup:2020ssz} to account for the large discrepancies among the four measures. The result determined in \cite{DAlise:2022ypp} yields: 
\begin{equation} \label{eq:AVG}
M_W^{\rm AVG} = 80409\pm  \, 17\,{\rm MeV}.
\end{equation}
The deviation from the SM reduces to around 3 standard deviations:
\begin{equation}
    \left.\Delta M_{\PW}\right|_{\rm AVG}= M_\PW^{\rm AVG} - M_\PW^{\rm th} =  52\pm 18~{\rm MeV}.
    \label{eq:AVG_with_CDF}
\end{equation}

\begin{table}
\begin{center}
\begin{tabular}{c|c|c|c|c}
  & LEP+TeVatron~\cite{ALEPH:2010aa} & ATLAS~\cite{ATLAS:2017rzl} & LHCb~\cite{LHCb:2021bjt} & CDF-II~\cite{CDF:2022hxs} \\ \hline
measurement & $80385\pm15~{\rm MeV}$ & $80370\pm19~{\rm MeV}$ & $80354\pm32{\rm MeV}$ & $80433.5\pm 9.4~{\rm MeV}$ \\
pull from SM & $28\pm16~{\rm MeV}$ & $13\pm 19~{\rm MeV}$ & $-3\pm 32~{\rm MeV}$ & $76\pm 11~{\rm MeV}$\\
\hline
\end{tabular}
\caption{\em\label{tab:MW} Recent measurements of the $\PW$  boson mass at colliders.}
\end{center}
\end{table}

\begin{figure}[tbh!]
    \centering
    \includegraphics[width = 0.7 \textwidth]{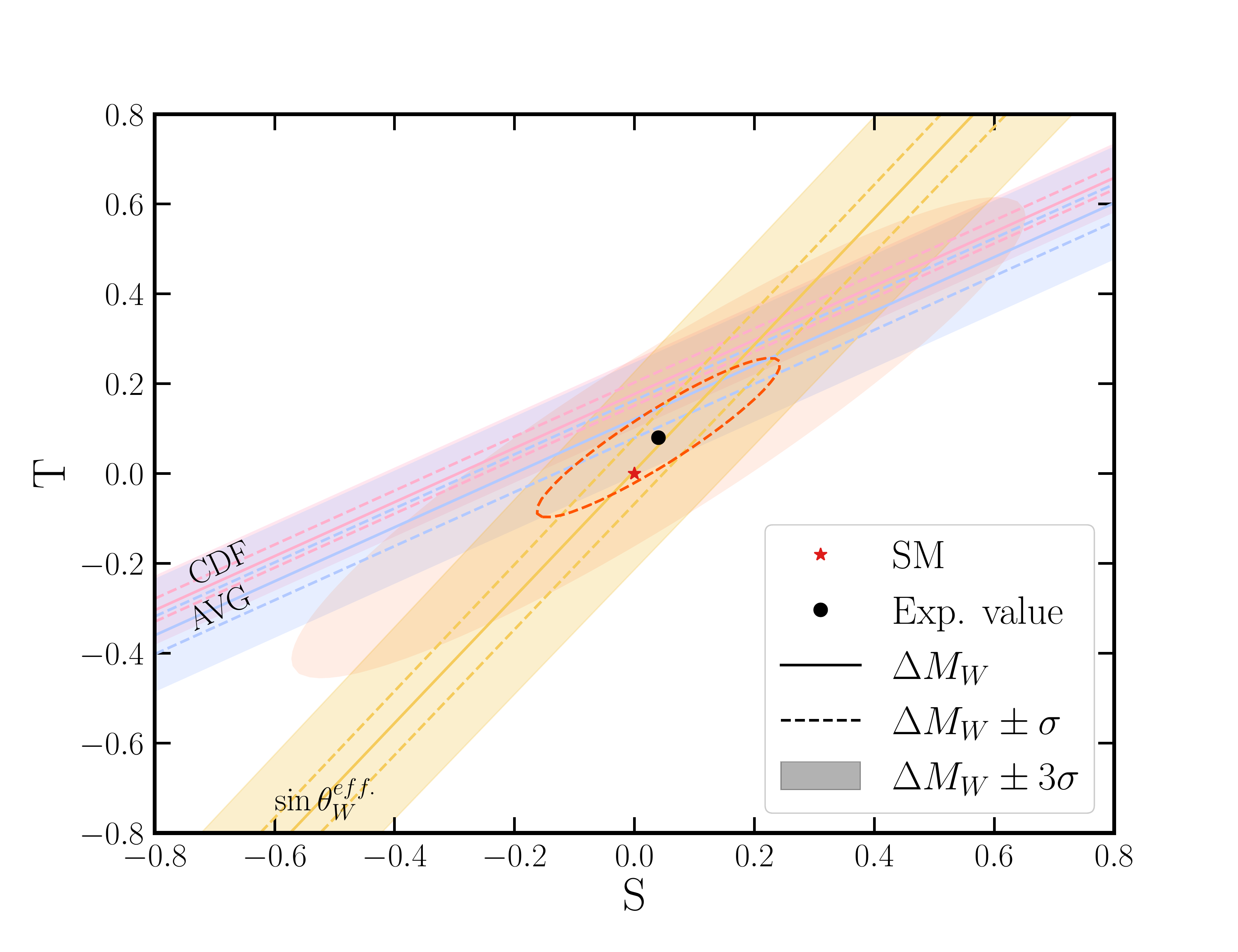}
    \caption{\em \label{fig:T_vs_S} In pink (violet) is shown the parameter space region that is able to explain the $\left.\Delta M_{\PW}\right|_{\rm CDF}$ ($\left.\Delta M_{\PW}\right|_{\rm AVG}$) within $3\sigma$. The black dot represents the experimental value of $T$ and $S$ and the orange ellipse is the corresponding covariance ellipse at $3\sigma$. The yellow region represents the bound originating from
   asymmetries and determination of $\sin^2 \theta_W^{eff}$.}
\end{figure}

It is possible to explain Eqs.\eqref{eq:CDF_W} or \eqref{eq:AVG} by introducing beyond the SM physics as summarized in  \cite{DAlise:2022ypp}. 
The corrections from NP to the $\PW$ boson mass can be expressed via the following approximate formula \cite{Altarelli:1994iz}:
\begin{equation}
    \Delta M_{\PW}^{\rm NP} \approx 300~{\rm MeV} \,\left(1.43\,T- 0.86\,S\right),
    \label{eq:delta M_W approx formula}
\end{equation}
where $T$ and $S$ are the so-called EW oblique parameters~\cite{Peskin:1990zt,Peskin:1991sw,Kennedy:1990ib}. We use the definition of $S$ and $T$ shown in Ref.~\cite{He:2001tp}, reported in the following:
\begin{equation}
   \begin{split}
       S & =  -16\pi\frac{\Pi_{3Y}(M_Z^2)-\Pi_{3Y}(0)}{M_Z^2},\\
       T & =  4\pi \frac{\Pi_{11}(0)-\Pi_{33}(0)}{s_W^2c_W^2M_Z^2},
   \end{split} 
   \label{eq:T and S general}
\end{equation}
where $s_W = \sin \theta_W$ and $c_W = \cos \theta_W$, $\theta_W$ being the Weinberg angle defined at the scale $\mu = M_Z$, $\Pi_{3Y}$ is the vacuum polarization of one ispospin and one hypercharge current, while $\Pi_{11}$ and $\Pi_{33}$ are the vacuum polarization of the isospin currents. 
We illustrate the impact of the $\PW$ boson mass measurement on EW physics in Fig.~\ref{fig:T_vs_S}, where a red star indicates the SM prediction at one-loop level, which corresponds to $T=S=0$. 
The black dot marks the observed value for the oblique parameters without including the CDF result, corresponding to $T=0.09\pm0.14$, $S=0.04\pm 0.11$, with a covariance of $\rho = 0.92$ \cite{Haller:2018nnx}. The dashed orange ellipse and the shaded orange ellipse correspond, respectively, to the covariance ellipse drawn at $1\sigma$ and $3\sigma$ of the oblique parameters measure. 
They include several EW precision measurements, including the $\PW$ boson mass: To separate its impact, we show the determination of $T$ and $S$ from separate sources. 
The light yellow dashed lines and the yellow region correspond to the bound originating from asymmetries measured at LEP experiments and from the determination of $\sin^2 \theta_W^{eff}$. The pink dashed line and the pink region is driven by the CDF mass measurement $\left.M_{\PW}\right|_{\rm CDF}$ within $1\sigma$ and $3\sigma$, respectively. 
The same reasoning applies to the light violet dashed lines and light violet region, which are relative to the average estimation $\left.M_{\PW}\right|_{\rm AVG}$. 
We remark that the new measurement pushes the preferred region toward positive values of $T$ with respect to the SM and the old results. This is confirmed by new EW fits that include the CDF result in the average \cite{Strumia:2022qkt,deBlas:2022hdk} or in replace of the old values \cite{Lu:2022bgw}. 
Hence, the new physics effect must be related to some breaking of the custodial symmetry defined in the SM Higgs sector.

\begin{figure}[tbh!]
\centering
\includegraphics[width = 0.5 \textwidth]{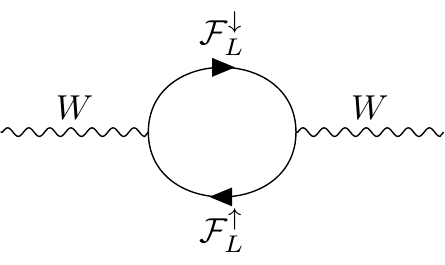}
\caption{\em Loop correction to $\PW$ boson mass stemming from the radiative model. \label{fig: loop diagram}}
\end{figure}

In the radiative model under investigation, this effect may dominantly arise from a mass difference between  the components of the doublet of fermions, $\FLu$ and $\FLd$, which contribute at one-loop level to the $\PW$ boson mass. In Fig.~\ref{fig: loop diagram}, we show the relevant diagram for this correction. 
This effect can be expressed in terms of the oblique parameters in Eq.\eqref{eq:T and S general}, which take the following expressions:
\begin{equation} \label{eq:SandT}
    \begin{split}
        S =\, & \frac{N_{TC}}{6\pi}\left\{2(4Y+3)x_1+2(-4Y+3)x_2-2Y\ln{\frac{x_1}{x_2}}+\right.\\
        &\left.+\left[\left(\frac{3}{2}+2Y\right)x_1+Y\right]g(x_1)+\left[\left(\frac{3}{2}-2Y\right)x_2-Y\right]g(x_2)\right\},\\
        T =\, &\frac{N_{TC}}{8\pi s_W^2c_W^2}f(x_1,x_2),\\
    \end{split}
\end{equation}
where $x_1= (m_{\FLu}/M_Z)^2$, $x_2= (m_{\FLd}/M_Z)^2$, and:
\begin{equation}
    \begin{split}
        f(x_1, x_2) & = \frac{x_1+x_2}{2}-\frac{x_1x_2}{x_1-x_2}\ln{\frac{x_1}{x_2}},\\
        g(x) & = -4 \sqrt{4x-1}\arctan\frac{1}{\sqrt{4x-1}}.
    \end{split}
\end{equation}
For equal masses, $x_1 = x_2$, the contribution to $T$ vanishes as $f(x, x) = 0$. This is clearly understood as such mass difference can only occur via breaking of the weak isospin, which is generated by the mixing induced by the couplings $\kup \neq \kdown$. Hence, the states running in the loop should be the final mass eigenstates after diagonalization. The results in Eq.\eqref{eq:SandT} are valid in the limit of small mixing. 

\begin{figure}[tbh!]
    \centering
    \includegraphics[width=0.49\textwidth]{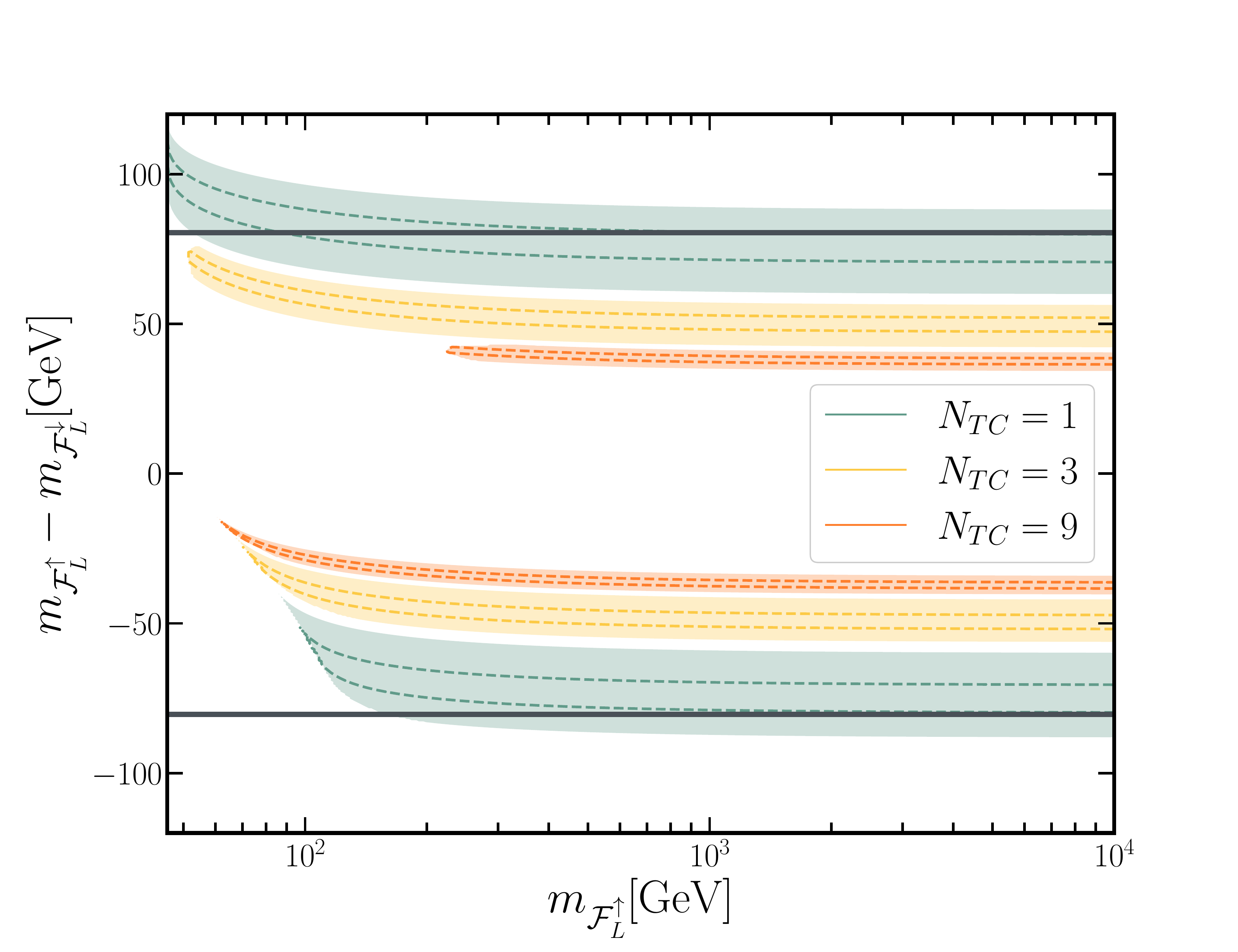}
      \includegraphics[width=0.49\textwidth]{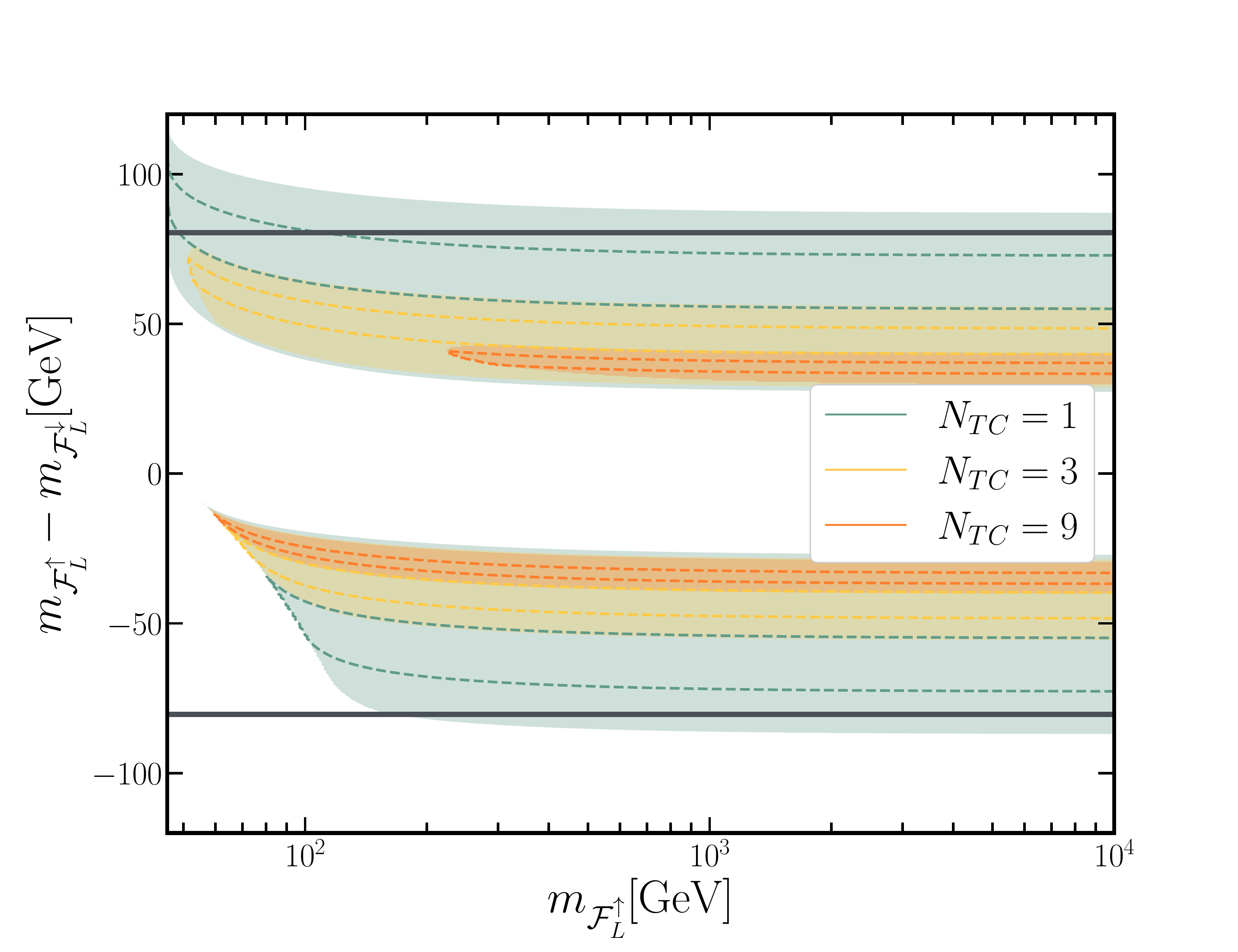}
    \caption{\em \label{fig:Delta M_W} In color is shown the parameter space that is able to compensate $\Delta M_{\PW}$ from CDF II (left panel) and the average value shown in Eq.\,\eqref{eq:AVG_with_CDF} (right panel). For each $N_{\rm TC}$, the bands represent the $3\sigma$ allowed region, while the dashed lines show the $1\sigma$ contour.}
\end{figure}

In Fig.~\ref{fig:Delta M_W} on the left (right) panel we show the parameter space region that is able to explain $\left.\Delta M_{\PW}\right|_{\rm CDF}$ ($\left.\Delta M_{\PW}\right|_{\rm AVG}$) within $3 \sigma$ for fixed values of $N_{TC}$, corresponding to different color shading. The dashed lines represent the agreement within $1\sigma$ with the experimental measurements. The plot investigates the dependence on the $\Fl$ mass versus the mass difference: The horizontal lines correspond to mass splits equal to $M_\PW$, reported as a reference. In the region above (below) the top (bottom) line, decays among the two components are kinematically open, hence allowing the on-shell decay
\begin{equation}
    \Fl^\uparrow\to \PW^- + \Fl^\downarrow\,\,\,\,\,\,\,\,\, \left( \Fl^\downarrow\to \PW^+ +\Fl^\uparrow\right)\,.
\end{equation}
The plots show that, for $N_{\rm TC} > 1$, the mass splits are required to be smaller than the $\PW$ mass, but still sizeable. Henceforth, the above decays can still occur via an off-shell $\PW$ boson.

If the isospin breaking in the fermion sector is too small, the leading effect may arise at two-loop level, involving the couplings $y_Q^{ij}$ with the third-generation quarks. The mass difference between top and bottom already induces a sizeable isospin violation, while the loop suppression can be compensated by the large $y_Q$ couplings of the third-generation quarks.

\section{Parameter space}\label{App:Parameter_space}
\begin{figure}
    \centering
    \includegraphics[width = 0.44 \textwidth]{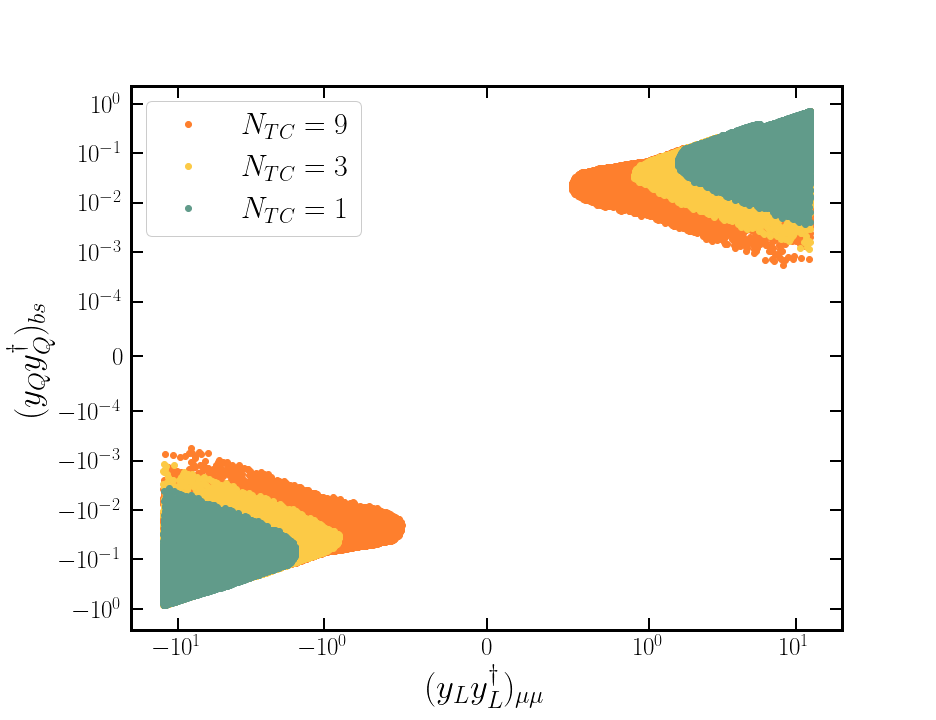}\\
    \includegraphics[width = 0.44 \textwidth]{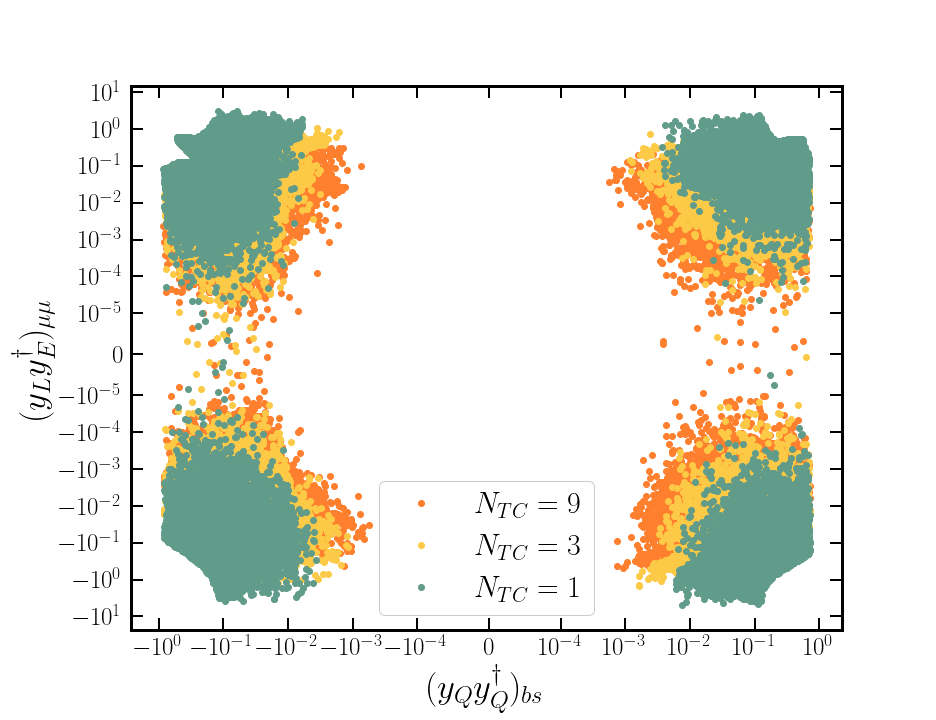}\,\,
    \includegraphics[width = 0.44 \textwidth]{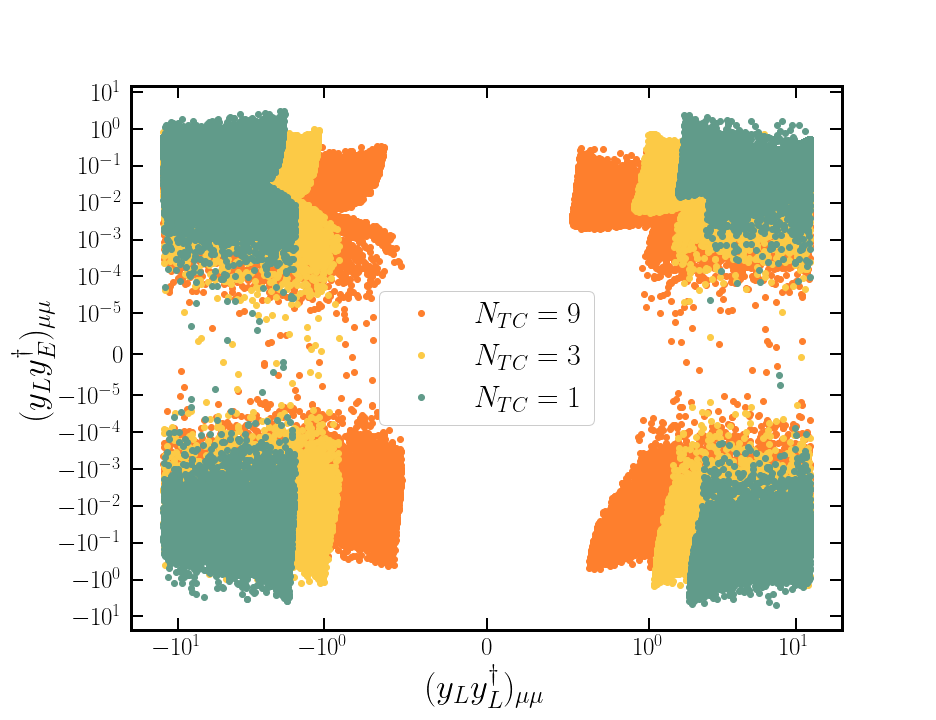}
    \caption{Parameter space for $(y_Qy_Q^{\dagger})_{bs}$ vs $(y_Ly_L^{\dagger})_{\mu\mu}$ (top), $(y_Ly_E^{\dagger})_{\mu\mu^K}$ vs $(y_Qy_Q^{\dagger})_{bs}$ (bottom left),and $(y_Ly_E^{\dagger})_{\mu\mu^K}$ vs $(y_Ly_L^{\dagger})_{\mu\mu}$ (bottom right) using the dispersive $\Delta a_\mu$ measurement~\cite{Abi:2021gix, Aoyama:2012wk,Aoyama:2019ryr,Czarnecki:2002nt,Gnendiger:2013pva,Davier:2017zfy,Keshavarzi:2018mgv,Colangelo:2018mtw,Hoferichter:2019mqg,Davier:2019can,Keshavarzi:2019abf,Kurz:2014wya,Melnikov:2003xd,Masjuan:2017tvw,Colangelo:2017fiz,Hoferichter:2018kwz,Gerardin:2019vio,Bijnens:2019ghy,Colangelo:2019uex,Blum:2019ugy,Colangelo:2014qya, Aoyama:2020ynm}.}
    \label{fig:Anomalies}
\end{figure}

\begin{figure}
    \centering
    \includegraphics[width = 0.44 \textwidth]{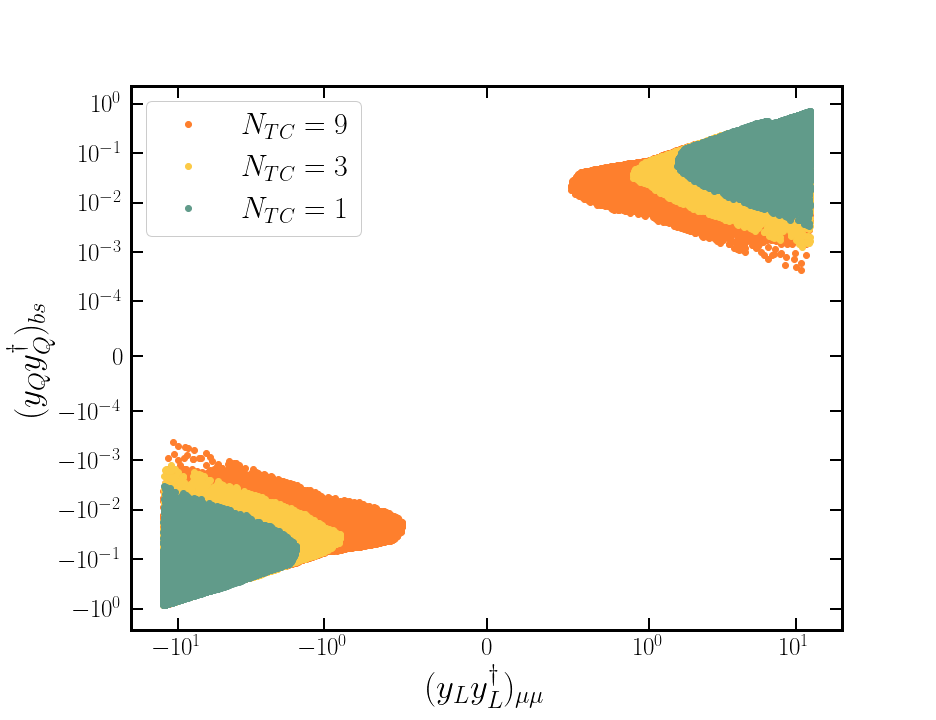}\\
    \includegraphics[width = 0.44 \textwidth]{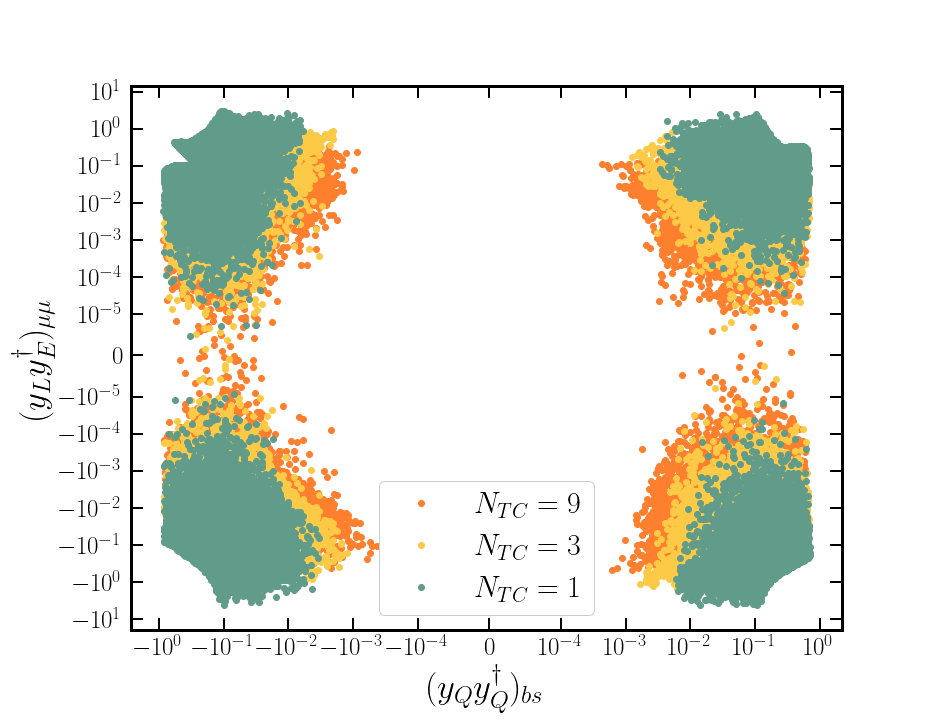}\,\,
    \includegraphics[width = 0.44 \textwidth]{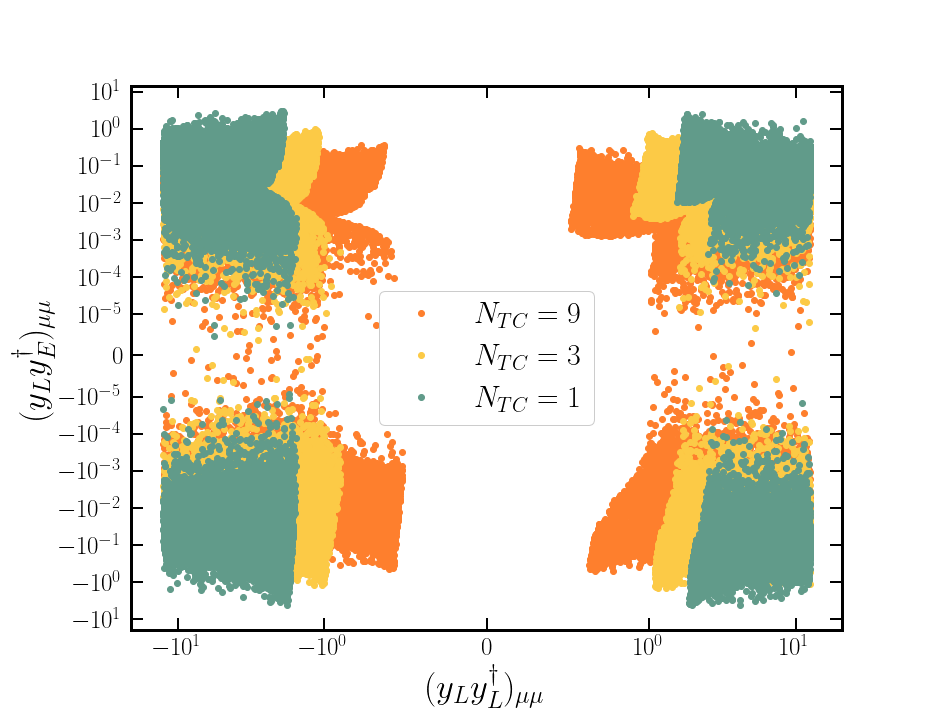}
    \caption{Parameter space for $(y_Qy_Q^{\dagger})_{bs}$ vs $(y_Ly_L^{\dagger})_{\mu\mu}$ (top), $(y_Ly_E^{\dagger})_{\mu\mu^K}$ vs $(y_Qy_Q^{\dagger})_{bs}$ (bottom left),and $(y_Ly_E^{\dagger})_{\mu\mu^K}$ vs $(y_Ly_L^{\dagger})_{\mu\mu}$ (bottom right) using the lattice $\Delta a_\mu$ measurement~\cite{Alexandrou:2022amy}.}
    \label{fig:Anomalies_lattice}
\end{figure}

Fig.s~\ref{fig:Anomalies} and~\ref{fig:Anomalies_lattice} show the available parameter space, respectively using the dispersive~\cite{Abi:2021gix, Aoyama:2012wk,Aoyama:2019ryr,Czarnecki:2002nt,Gnendiger:2013pva,Davier:2017zfy,Keshavarzi:2018mgv,Colangelo:2018mtw,Hoferichter:2019mqg,Davier:2019can,Keshavarzi:2019abf,Kurz:2014wya,Melnikov:2003xd,Masjuan:2017tvw,Colangelo:2017fiz,Hoferichter:2018kwz,Gerardin:2019vio,Bijnens:2019ghy,Colangelo:2019uex,Blum:2019ugy,Colangelo:2014qya, Aoyama:2020ynm} and lattice~\cite{Alexandrou:2022amy} $\Delta a_\mu$ measurements, in the planes $(y_Qy_Q^{\dagger})_{bs}$ vs $(y_Ly_L^{\dagger})_{\mu\mu}$, $(y_Ly_E^{\dagger})_{\mu\mu^K}$ vs $(y_Qy_Q^{\dagger})_{bs}$,and $(y_Ly_E^{\dagger})_{\mu\mu^K}$ vs $(y_Ly_L^{\dagger})_{\mu\mu}$.

\section{Loop Functions}\label{App:Loop}
In this section we report the Loop functions used in \ref{sec:prec_constraints}.
\begin{equation}
\begin{split}
    F(x,y)&=\frac{1}{(1-x)(1-y)}+\frac{x^2\ln x}{(1-x)^2(x-y)}+\frac{y^2\ln y}{(1-y)^2(y-x)}\,,\\
     F(x,1)&=F(1,x) =\frac{-1+4x-3x^2+2x^2\ln x}{2(-1+x)^3}\,,\\
    F(x,x)& = \frac{1-x^2+x\ln x}{(1-x)^3} \,,\\
    F(1,1)&=\frac{1}{3}\,.
    \end{split}
    \label{eq:F(x,y)}
\end{equation}

\begin{equation}
    \begin{split}
    \tilde{F}_7 (y) &= \frac{F_7(y^{-1})}{y}= \frac{1-6y+ 3y^2+2y^3+6y^2\ln y}{12(1-y)^4}\,,\\
       \tilde{F}_{7}(1) & = \frac{1}{24}\,,
    \end{split}
\end{equation}
\begin{equation}
\begin{split}
     G_{LR}(y) &= \frac{1-4y+3y^2-2y^2\ln y}{2(1-y)^3}\,,\\
      G_{LR}(1) & = \frac{1}{3}\,,
\end{split}
\end{equation}
\begin{equation}
    \begin{split}
        F_{LR}(y) &=1\,,\\
        F_{LR}(1) &=\frac{1}{6}\,.
    \end{split}
\end{equation}

Such functions were taken from \cite{Arnan:2016cpy, Calibbi:2018rzv} and adapted to our case.
\newpage 
\section{Diagrams}\label{App:Diagrams}
\subsection{Single $\SD$}
\begin{figure}[tbh!]
    \centering
    \includegraphics[height = 0.2 \textwidth]{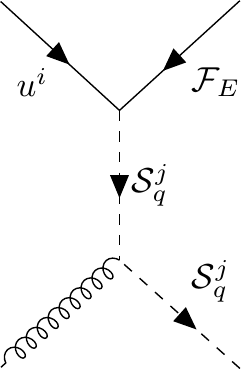}\,\,\,
    \includegraphics[height = 0.2 \textwidth]{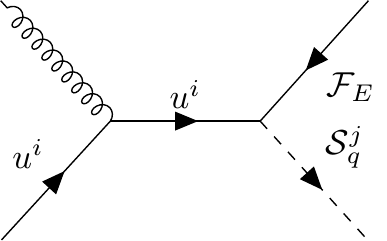}\,\,\,
    \includegraphics[height = 0.2 \textwidth]{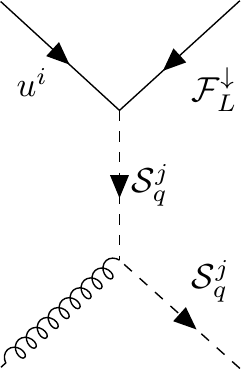}\,\,\,
    \includegraphics[height = 0.2 \textwidth]{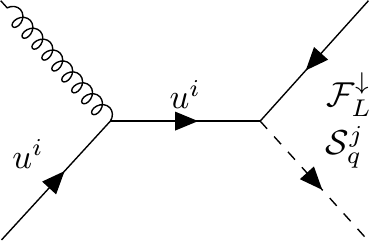}
   \label{fig:production_pair_app}
\end{figure}
\begin{figure}[tbh!]
    \centering
    \includegraphics[height = 0.2 \textwidth]{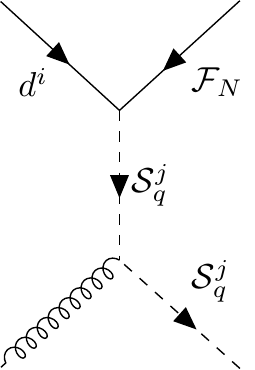}\,\,\,
    \includegraphics[height = 0.2 \textwidth]{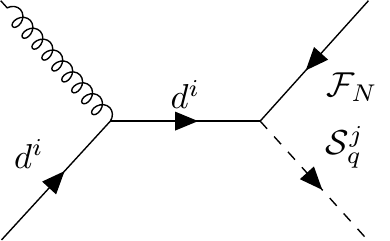}\,\,\,
    \includegraphics[height = 0.2 \textwidth]{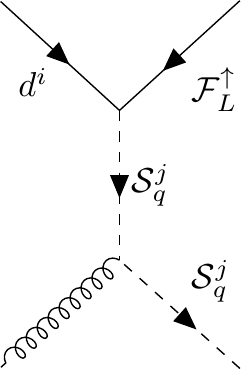}\,\,\,
    \includegraphics[height = 0.2 \textwidth]{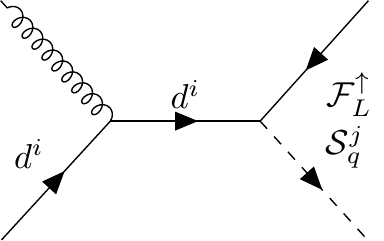}
   \label{fig:gd}
\end{figure}
\subsection{Double $\F$}
\begin{figure}[tbh!]
    \centering
    \includegraphics[height = 0.2 \textwidth]{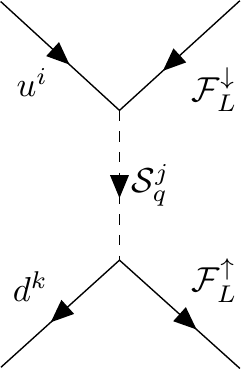}\,\,\,
    \includegraphics[height = 0.2 \textwidth]{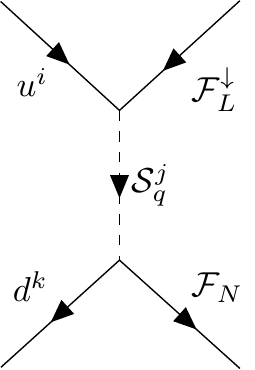}\,\,\,
    \includegraphics[height = 0.2 \textwidth]{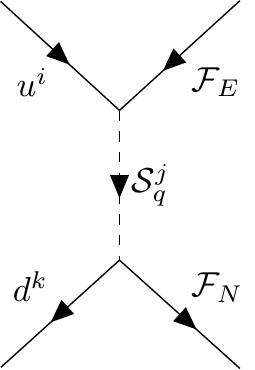}\,\,\,
    \includegraphics[height = 0.2 \textwidth]{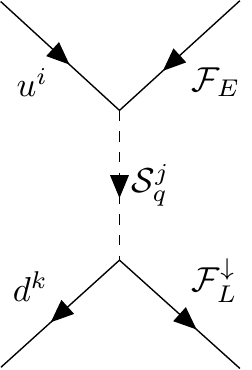}
   \label{fig:ud}
\end{figure}
\begin{figure}[tbh!]
    \centering
    \includegraphics[height = 0.2 \textwidth]{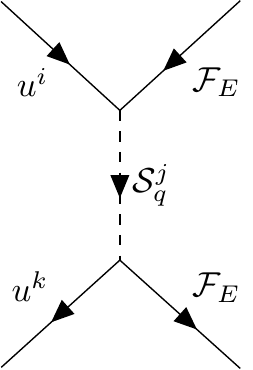}\,\,\,
    \includegraphics[height = 0.2 \textwidth]{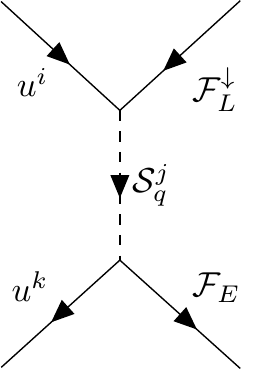}\,\,\,
    \includegraphics[height = 0.2 \textwidth]{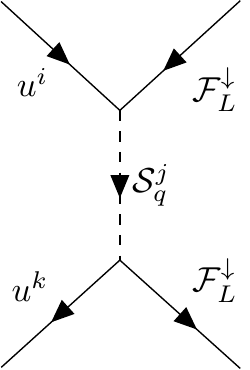}\,\,\,
    \includegraphics[height = 0.2 \textwidth]{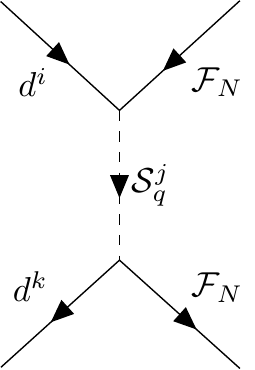}\,\,\,
    \includegraphics[height = 0.2 \textwidth]{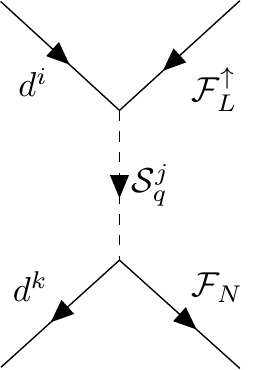}\,\,\,
    \includegraphics[height = 0.2 \textwidth]{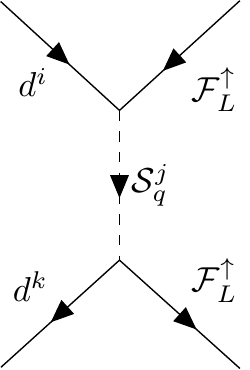}
   \label{fig:uu and dd}
\end{figure}
\begin{figure}[tbh!]
    \centering
    
   \label{fig:dd}
\end{figure}

\end{document}